\documentclass[lettersize,journal]{IEEEtran}
\usepackage{amsmath,amsfonts}
\usepackage{algorithmic}
\usepackage{algorithm}
\usepackage{array}
\usepackage[caption=false,font=normalsize,labelfont=sf,textfont=sf]{subfig}
\usepackage{textcomp}
\usepackage{stfloats}
\usepackage{url}
\usepackage{verbatim}
\usepackage{graphicx}
\usepackage{cite}

\usepackage{booktabs}

\begin{document}

\title{
Lightweight Fingernail Haptic Device: Unobstructed Fingerpad Force and Vibration Feedback for Enhanced Virtual Dexterous Manipulation}

\author{Yunxiu Xu,~\IEEEmembership{Student Member,~IEEE}, Siyu Wang, and Shoichi Hasegawa
\thanks{This work was supported by JST, the establishment of university fellowships towards the creation of science and technology innovation, Grant Number JPMJSP2180, and JSPS KAKENHI Grant Number 23H03432.

Y. XU, S. Wang and S. Hasegawa are with the Department of Information
and Communications Engineering, School of Engineering, Institute of Science Tokyo, Japan, e-mail: \{yunxiu, siw131, hase\}@haselab.net}}

\markboth{IEEE TRANSACTIONS ON HAPTICS,~Vol.~14, No.~8, August~2021}%
{Shell \MakeLowercase{\textit{et al.}}: A Sample Article Using IEEEtran.cls for IEEE Journals}

\IEEEpubid{0000--0000/00\$00.00~\copyright~2021 IEEE}

\maketitle

\vspace{0.5em}
\noindent \textbf{Copyright Notice:} © 2025 IEEE. Personal use of this material is permitted. Permission from IEEE must be obtained for all other uses, in any current or future media, including reprinting/republishing this material for advertising or promotional purposes, creating new collective works, for resale or redistribution to servers or lists, or reuse of any copyrighted component of this work in other works.
\vspace{0.5em}

\begin{abstract}
This study presents a lightweight, wearable fingertip haptic device that provides physics-based haptic feedback for dexterous manipulation in virtual environments without hindering real-world interactions. 
The device, designed with thin strings and actuators attached to the fingernails, ensures minimal weight (1.55 g per finger) and preserves finger flexibility. Integrating the software with a physics engine renders multiple types of haptic feedback (grip force, collision, and sliding vibration feedback).
We evaluated the device's performance in pressure perception, slip feedback, typical dexterous manipulation tasks, and daily operations, and we gathered user experience through subjective assessments. Our results show that participants could perceive and respond to pressure and vibration feedback. Through dexterous manipulation experiments, we further demonstrated that these minimal haptic cues significantly improved virtual task efficiency, showcasing how lightweight haptic feedback can enhance manipulation performance without complex mechanisms. The device's ability to preserve tactile sensations and minimize hindrance to real-world operations is a key advantage over glove-type haptic devices. This research offers a potential solution for designing haptic interfaces that balance lightweight construction, haptic feedback for dexterous manipulation, and daily wearability.

\end{abstract}

\begin{IEEEkeywords}
Virtual reality, lightweight haptic display, tactile devices, dexterous manipulation, wearable haptics
\end{IEEEkeywords}

\section{Introduction}
\IEEEPARstart{V}{irtual} Reality (VR) technology has continuously pushed beyond traditional physical space limitations. This progress inspires expectations for extended periods of work, creation, and social interaction in virtual spaces. VR has the potential to redefine our ``daily life" by offering users novel experiences and interaction methods. However, a critical challenge persists in current virtual environments: the lack of effective haptic feedback for hands.
Since haptic feedback plays a crucial role in coordinating visual and physical perceptions, its absence leads to a sensory information discrepancy, affecting user immersion, satisfaction, and operational accuracy \cite{macfarlane1999force}. Without haptic feedback, users must rely mostly on visual and auditory information to perform tasks \cite{zangrandi2021neurophysiology}.
This issue becomes particularly urgent as researchers and engineers work to create virtual environments that accurately reproduce multi-sensory experiences. Understanding and replicating the tactile mechanisms of human interaction with the world has become a research focus.
The realism of VR experiences depends on the quality of interactions between the virtual hand and virtual objects \cite{mangalam2024enhancing}. How to accurately reproduce haptic stimuli on users' hands to achieve precise physical control has been a significant direction in haptic interaction research. Achieving this goal not only helps improve the precision and reliability of complex tasks (such as precision assembly) but also provides users with more natural and intuitive interaction experiences.

\begin{figure}[!t]
\centering
\includegraphics[width=2.5in]{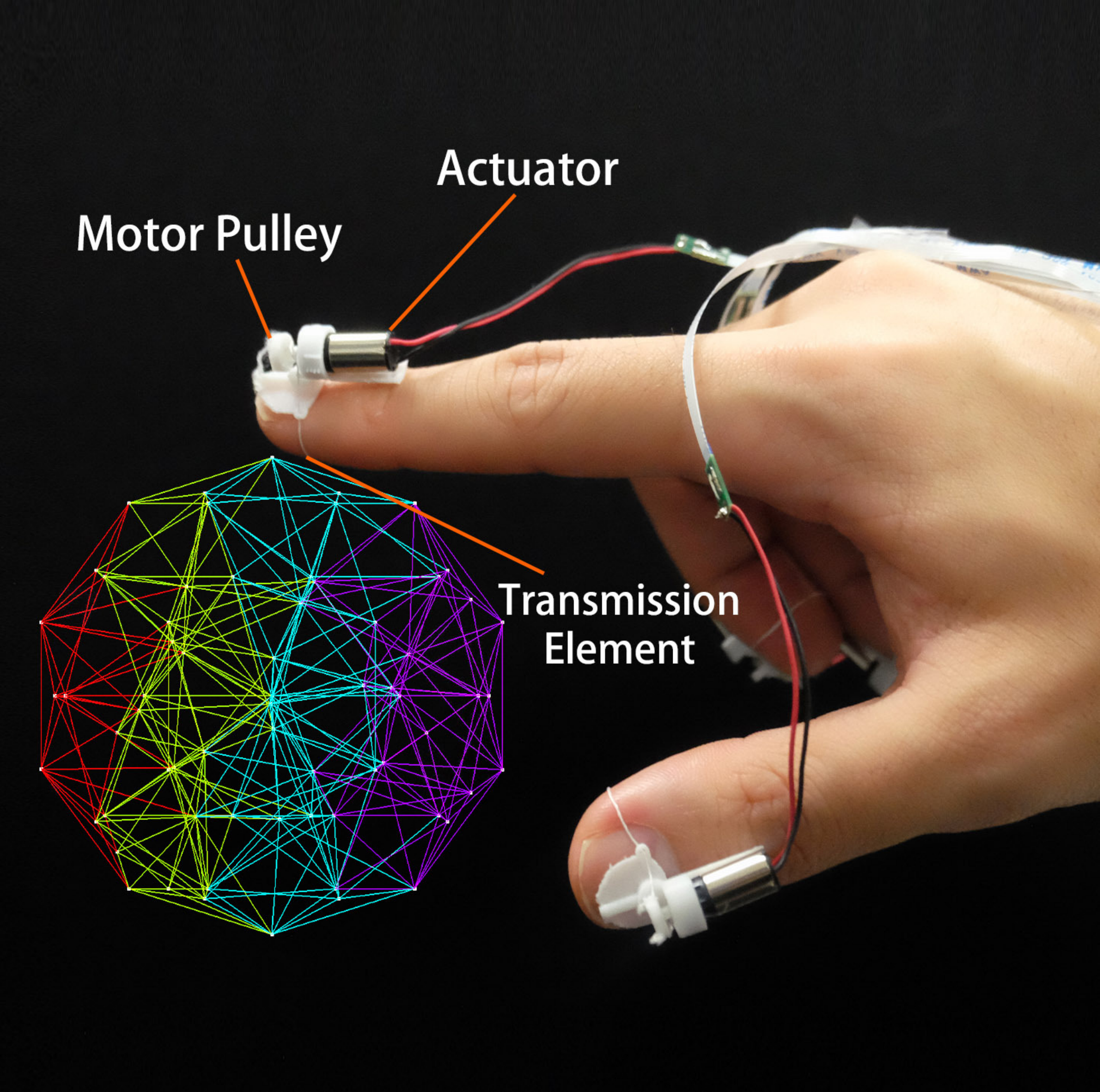}
\caption{The fingertip haptic device mounts on the fingernail, providing haptic feedback on the fingerpad side through a thin string driven by a motor. This design keeps the pad of the fingertip unobstructed, allowing the user to feel real objects while receiving haptic sensations. The device weighs 1.55 g per finger.}
\label{devicePhoto}
\end{figure}

\IEEEpubidadjcol

This paper proposes a nail-mounted device that applies normal force and vibration to the fingerpad via a string to facilitate dexterous manipulation in XR (Extended Reality, including VR, AR, and MR) by providing grip force, contact, and sliding feedback through tension, shown in Fig. \ref{devicePhoto}. It does not interfere with other fingers and real object manipulation.

To achieve effective virtual object manipulation, our proposed system addresses three key issues:
First, it provides essential haptic cues required for manipulation: contact feedback, sliding vibration feedback, and grip force feedback through a simple string mechanism.
Second, its small and lightweight design ensures it does not interfere with natural finger movements during virtual manipulation tasks.
Third, it delivers feedback to the fingertip without covering the fingerpad, allowing for both realistic virtual feedback and unobstructed manipulation of real objects during long-time daily wear.

The manuscript is structured as follows. Section I introduces the motivation and challenges of providing effective haptic feedback in virtual reality and reviews related work. Section II presents the proposed system design and implementation. Section III evaluates the haptic device's performance. Section IV discusses limitations and potential improvements, while Section V concludes with key findings and future research directions. In the following parts, we review related work in haptic devices and tactile feedback.

\subsection{The Principles Behind Dexterous Haptic Perception}
Dexterous manipulation in VR is an area where multiple human fingers, with their rich mechanoreceptors in palms and fingertips, cooperate to grasp and manipulate virtual objects \cite{okamura2000overview}. Dexterous manipulation involves determining the motion state of the object, the contact state between the object and the fingers, and the friction characteristics. Some studies have explored the control principles of the human brain for dexterous manipulation. Previous research \cite{johansson2009coding} demonstrated that automatic adjustment of the grip force occurs when an object undergoes unexpected acceleration due to instantaneous sliding between the fingertips and the object. Zangrandi et al. \cite{zangrandi2021neurophysiology} have further elucidated the neurophysiological basis of slip sensation and grip reaction, showing that slippage control involves parallel activation of multiple mechanoreceptors. Wiertlewski et al. \cite{Wiertlewski2013Slip} found that the central nervous system can perceive the movement state of an object through vibrations generated by the friction between the skin and the object's surface, and adjust the grip force to terminate sliding.

Based on these studies of human manipulation mechanisms, we designed our haptic feedback system to support the dexterous manipulation of objects in virtual environments. We then evaluated how our device impacts users' abilities.
In this research, we utilize the simulation of different skin mechanoreceptors, using continuous pressure to activate SA1 receptors and vibrations to stimulate FA1 and FA2 receptors to provide natural and precise haptic feedback for virtual object manipulation.

\subsection{Traditional Haptic Devices Designed for Manipulation}

Various haptic devices provide feedback to the fingers, such as haptic gloves, hand-held devices, and devices that control finger activity through locking mechanisms \cite{haptx2022, hinchet2018dextres, pacchierotti2017wearable, adilkhanov2022haptic}. 
However, these devices control hand movements by applying external forces to impose mechanical constraints, requiring a stable support point like the wrist, arm, or shoulder to withstand the reaction force. The overall volume of these solutions is difficult to reduce, and they decrease hand comfort and freedom. 
But these design compromises are what enable the simultaneous delivery of realistic tactile and force feedback. Many studies eliminate force feedback by focusing only on fingertip tactile feedback, omitting whole-hand mechanisms. Schorr et al. \cite{schorr2017fingertip} developed a device for skin deformation on fingerpads. Also, several 3-degree-of-freedom (DOF) and 2-DOF \cite{prattichizzo2013towards, pacchierotti2017wearable, adilkhanov2022haptic} devices were proposed. However, these devices require bulky mechanical structures on the fingertip, which place a burden on the hands, impair precise finger movements, and prevent natural interaction with real objects.

\subsection{Emerging Haptic Devices Aimed at Preserving Real-world Manual Dexterity}

Recent research has explored various approaches to preserve natural tactile feedback during haptic interactions. One category of solutions focuses on relocating the actuators away from the fingerpad, such as placing them on fingernails, finger sides, proximal phalanx, or using skin stretch at the wrist for grasping by fingers \cite{preechayasomboon2021haplets, adilkhanov2022haptic}. Although these methods can preserve the fingerpad's tactile sensation and are easy to wear, they cannot stimulate the fingerpads directly.
Another approach uses mechanical designs that can temporarily move palmar components away when not needed \cite{teng2021touch}. An additional category eliminates body-worn actuators entirely by using contactless stimulation methods like ultrasound \cite{Matsubayashi2019Direct} or air pressure \cite{Gupta2013Airwave}. Alternative approaches include electrostimulation techniques \cite{tanaka2023full, tanaka2024reawristic} and material-based solutions like thin films \cite{withana2018tacttoo}, electroosmotic devices \cite{shen2023fluid}, and hydraulic/electrohydraulic systems \cite{feng2017submerged, hartcher2023fingertip}.

While these approaches offer potential advantages in preserving fingerpad tactile sensation, they face several common challenges. Mechanical relocation methods often introduce increased latency and learning time, while making the devices bulky and difficult to mount. Contactless stimulation requires unobstructed paths between actuators and users, limiting practical applications. 
While some electrotactile devices can stimulate fingertip tactile receptors without direct attachment to the fingerpad, they face challenges in selectively activating deeper mechanoreceptors without co-stimulating shallower ones \cite{kajimoto2004electro}. This limitation affects their ability to accurately reproduce complex mechanical sensations like pressure gradients. In contrast, mechanical approaches can more precisely target specific tactile sensations, though they often require direct contact with the fingerpad, which our design aims to avoid.
Material-based solutions struggle with either durability limitations or require bulky supporting components that cover significant portions of the fingerpad.

These limitations highlight the need for a haptic interface that can provide effective tactile feedback while maintaining the natural sensitivity and dexterity of the fingerpads. Our proposed approach addresses these challenges through a lightweight string-based mechanism that preserves fingerpad accessibility while enabling precise haptic stimulation.


\subsection{Relationship to Prior Work}
Our current work evolves from several foundational studies.
Aoki et al. \cite{aoki2009wearable} developed a string-based fingerpad device using micro electromagnets, which had a maximum output force of 420 mN. However, their design could not sustain continuous maximum power output without rapid burnout, which limited their ability to provide consistent pressure feedback. Although their approach uses strings to deliver fingertip feedback, our implementation focuses on enabling dexterous manipulation. The proposed device employs an alternative actuation mechanism capable of delivering a continuous tension output of up to 1.36 N, whereas Aoki's approach lacked independent control over tension and vibration. Additionally, the proposed method improves the rendering of grip contact and sliding sensations, both of which are crucial for grip simulation. The proposed device also extends from a single-finger setup to a multi-finger implementation, a necessary advancement for manipulation.
Compared to our previous demonstration \cite{xu2023realistic}, this paper removes the ring structure that occupied the palmar side. Following the experimental design in \cite{xu2024optimizing}, we conducted experiments evaluating vibrotactile and pressure feedback, including assessments of user experience in practical manipulation tasks. These evaluations, utilizing quantitative metrics, extended our understanding of this string-based system and demonstrated its contribution to virtual reality manipulation interactions.
Our haptic rendering approach specifically addresses grip contact and sliding feedback, which represents another important difference from previous work. Table \ref{tab:comparison} provides a detailed comparison of our device with several related wearable haptic systems. 

\begin{table*}[!t]
  \centering
  \caption{Comparison of Proposed Device with Related Wearable Haptic Systems}
  \label{tab:comparison}
  \resizebox{\textwidth}{!}{
  \begin{tabular}{@{}l p{3cm} p{2.5cm} p{2.5cm} p{2.5cm} p{3cm}@{}} 
    \toprule
    \textbf{Feature} & \textbf{This Work} & \textbf{Gravity Grabber} \cite{Minamizawa2007Gravity} & \textbf{Haplets} \cite{preechayasomboon2021haplets} & \textbf{Fingeret} \cite{maeda2022fingeret} & \textbf{Fluid Reality} \cite{shen2023fluid} \\
    \midrule
    Fingerpad Obstruction & \textbf{No} & Yes & \textbf{No} & \textbf{No} & Yes\\
    \midrule
    Feedback Point & Fingerpad (string) & Fingerpad (belt) & Finger Dorsum/Nail & Fingertip Sides \& Fingernail & Fingerpad\\
    \midrule
    Actuation Mechanism & Single DC Motor + String & Dual DC Motors + Belt & Linear Resonant Actuator (LRA) & Dual DC Motors + LRA & Electroosmotic Pump Array \\
    \midrule
    Feedback Type(s) & Normal Force, Vibration & \textbf{Normal Force, Shear Force, Vibration} & Vibration & Skin deformation \& Vibration & \textbf{Shape/Deformation} \\
    \midrule
    Spatial Resolution & Line Contact & Surface Contact & Single Point & Line Contact & \textbf{High (20 pixels/cm²)} \\
    \midrule
    Weight (On Finger) & \textbf{1.55 g} & Tens of grams & 5.2 g & 18 g & 6.2 g \\
    \bottomrule
  \end{tabular}%
  } 
\end{table*}

\section{System Overview}

\begin{figure}[!t]
  \centering
  \includegraphics[width=3.1in]{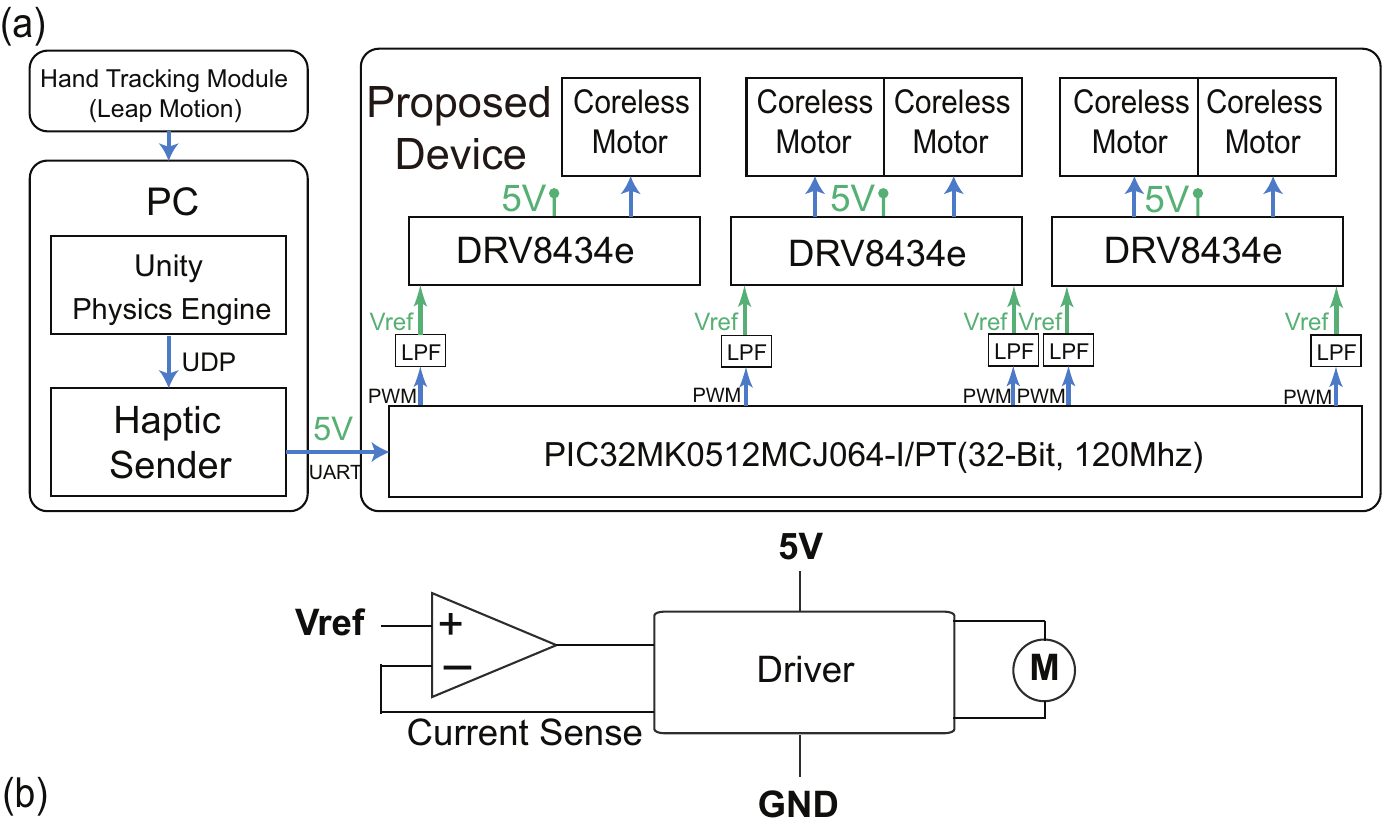}
  \caption{(a) The schematic diagram illustrates the information flow and main components of the system. (b) Control diagram of the motor-current feedback control. All components except the motor are embedded inside the motor driver chip.
  }
  \label{systemFigure}
  \end{figure}

The proposed haptic feedback system integrates components to help dexterous manipulation, including grip and slide, in virtual environments. As shown in Fig. \ref{systemFigure}, our implementation consists of the fingernail-mounted mechanical unit connected to the control circuit board that processes signals from a physics engine to generate appropriate haptic responses.

\begin{figure}[!t]
  \centering
  \includegraphics[width=3.3in]{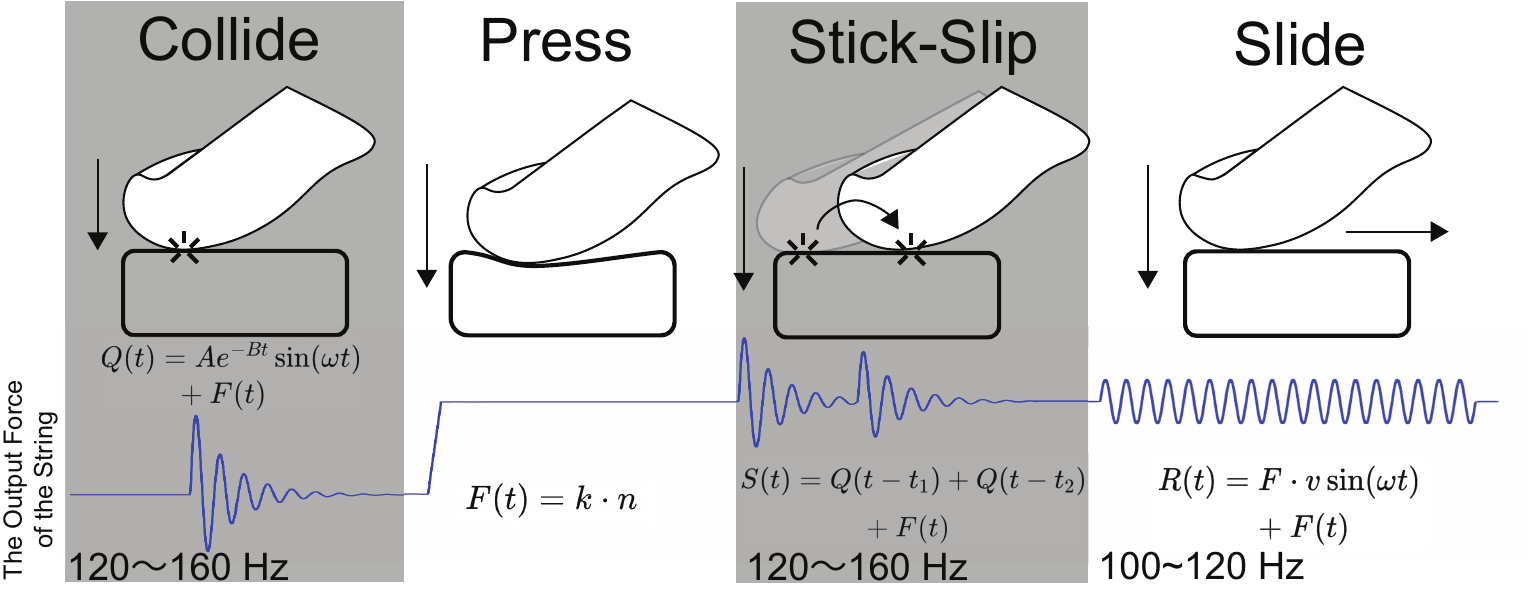}
  \caption{The figure illustrates the design of four key interaction modes between a fingertip and an object, along with their corresponding haptic feedback signals. The top row shows the physical interactions: Collide, Press, Stick-slip, and Slide. }
  \label{feedbackimg}
  \end{figure}

\subsection{Core Proposal}

\subsubsection{Haptic Feedback Concept}

Our haptic feedback concept, illustrated in Fig. \ref{feedbackimg}, addresses four fundamental interaction modes between fingertips and objects: Collide (impact), Press (continuous force), Slide (sliding vibration, without tangential force and directional information), and Stick-slip (sitck-slip vibration, transition between static and kinetic friction, accompanied by sliding vibration). Each mode is mapped to specific haptic signals designed to stimulate target mechanoreceptors, with Collide feedback and Stick-slip targeting FA2 receptors and Slide targeting FA1 receptors, while Press feedback stimulates SA1 receptors. The system does not provide tangential force feedback, focusing instead on normal forces and vibrotactile stimulation.

\subsubsection{Mechanical Design Principles}
Our design followed several key principles to achieve effective haptic feedback while maintaining wearability. For the fingertip unit, we prioritized minimizing weight and size while maximizing force output. The device mounts on the fingernail, keeping the fingerpad unobstructed for natural tactile sensation. 
The string-based mechanism was chosen for its ability to deliver normal force to the fingerpad without covering it. For the string-winding axle design, a smaller axle radius is preferable since the string tension force is inversely proportional to the axle radius. The motor's own axle can be used without additional axle designs.

\subsubsection{Physics-Based Hand Model}
\begin{figure}[!t]
  \centering
  \includegraphics[width=3.1in]{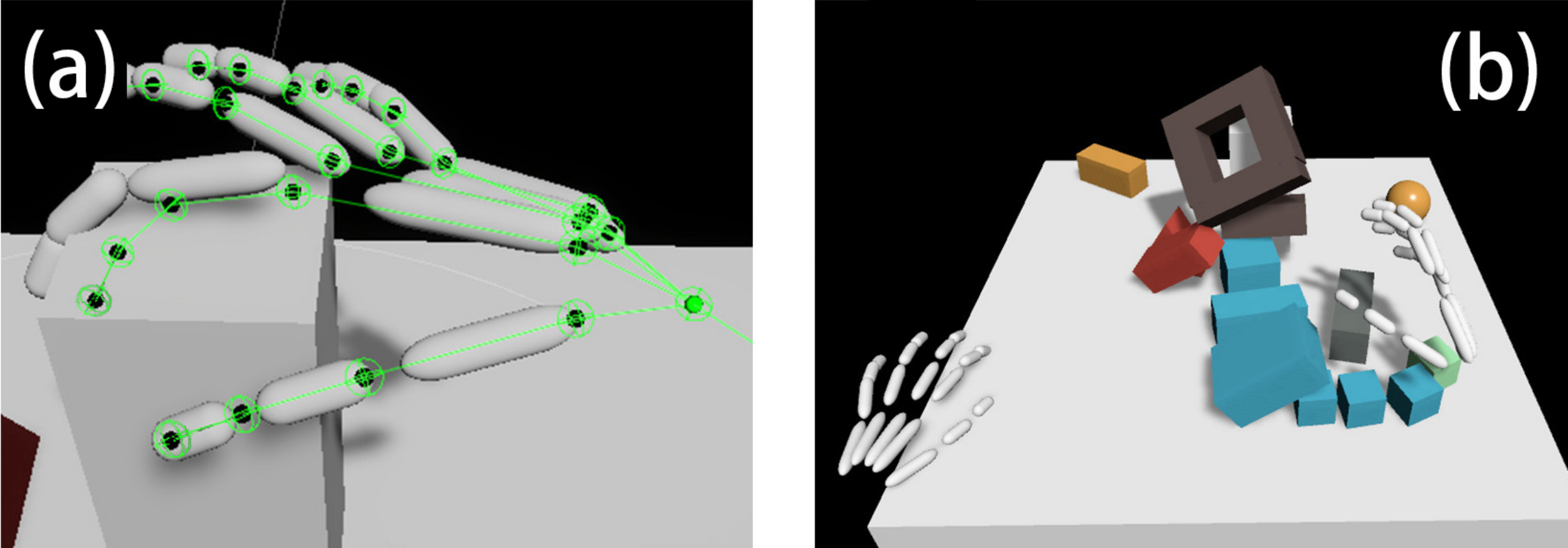}
  \caption{(a) The virtual hand is composed of virtual couplings, with the force output proportional to the penetration depth of the real hand's position (in green). The phalanges measured from the real hand and their corresponding simulated counterparts in the physics engine are connected by virtual springs. Also, adjacent phalanges are connected by ball joints. (b) A sample physics-based dexterous manipulation scene where participants can perform tasks such as stacking objects, placing objects in holes, and sliding objects.}
  \label{UnityScene}
  \end{figure}

The physics simulation for the virtual hand in our haptic feedback system integrates methods that ensure stable performance. In contrast to the complex spring model of torsional and linear springs employed by Borst et al. \cite{borst2006spring}, we directly couple each tracked phalange with its simulated counterpart using a spring-damper system. To further enhance the model, adjacent phalanges are connected via ball joints, as illustrated in Fig. \ref{UnityScene}(a). 

\subsection{Technical Realization}
\begin{figure}[!t]
  \centering
  \includegraphics[width=2.0in]{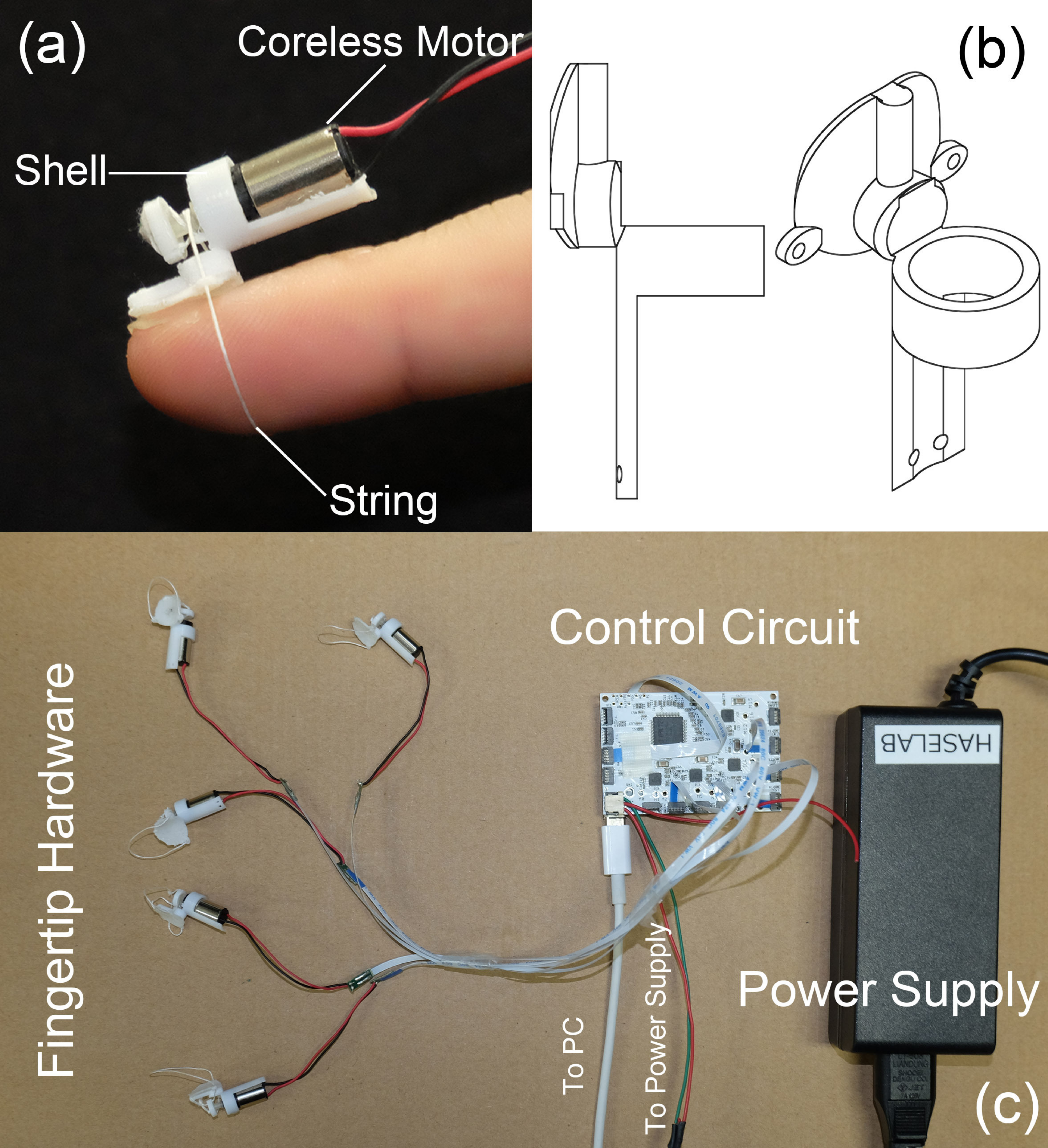}
  \caption{Hardware implementation of the proposed device. (a) The fingertip hardware components show the motor, string, and shell structure mounted on the fingernail using silicone adhesive. (b) The device shell, with holes that guide the string path to prevent accidental over-winding around the motor axle. (c) The control circuit board integrates the microcontroller, motor drivers, and supporting components, which are worn on the wrist. The fingertip unit and control board are lightweight, making the system comfortable to wear. The system is powered by a power supply.}
  \label{smallDevicePhoto}
  \end{figure}
\subsubsection{Hardware Implementation}
As shown in Fig. \ref{systemFigure} and Fig. \ref{smallDevicePhoto}, our hardware consisted of a PIC32MK0512MCJ064-I/PT microcontroller (Microchip Technology Inc., USA), capable of outputting up to 9 PWM channels. 
The DRV8434e motor driver (Texas Instruments Inc., USA) utilizes Smart Tune Ripple Control mode to control the motor current, for which PWM signals are converted into the current control voltages by RC low-pass filters.
The motor we used was a 6 mm x 14 mm DC Coreless Planetary Gear Motor with a gear ratio of 5.14 (Generic, China) as shown in Fig. \ref{measureSet}(a).
The string was made of ultra-high-molecular-weight polyethylene and has a diameter of 0.26 mm. For attachment to the fingernail, we applied BBX909 acrylic-modified silicone pressure-sensitive adhesive (CEMEDINE Co., Ltd., Japan) to the device's contact surface. After a 24-hour curing period, the adhesive allows for repeated use. Before mounting, both the fingernail and the device's adhesive area were cleaned with 75\% alcohol. This adhesive formed a bond under light pressure, permitted residue-free removal, retained bonding strength after repeated use, and left no residue after wearing. The power supply (LI TONE ELECTRONICSCO., LTD) provided an output of 5 V.

\subsubsection{Software Implementation}
For the software component, we utilized Unity (Unity Technologies, USA) to create virtual scenes and visual effects shown in Fig. \ref{UnityScene}(b). For physics engine integration and high-quality haptic rendering, our haptic feedback system used Springhead \cite{Shoichi2012springhead}. The haptic feedback system leveraged Springhead's capabilities to provide richer physical interaction data compared to Unity's built-in physics, including contact force between rigid bodies and dynamic or static friction states. For collision handling, the proposed system combines a ray-casting approach, employing the Gilbert-Johnson-Keerthi (GJK) algorithm for continuous collision detection \cite{factory2004ray}, with a Linear Complementarity Problem (LCP)-based method to compute collision response velocities. Furthermore, we adopted the linearly implicit time-stepping scheme proposed by Anitescu et al. \cite{anitescu2002time}, which efficiently manages the stiff dynamics involving contact, friction, and virtual coupling. This stable approach allows us to operate at a relatively low update rate (100 Hz) without sacrificing the accuracy or stability typically compromised in explicit methods.

The following describes the implementations of haptic signals:
\begin{itemize}
\item Based on \cite{okamura1998vibration}, $Q(t) = Ae^{-Bt}\sin(\omega t) + F(t)$ represents the vibration signal generated during the collision, aiming at stimulating FA2 receptors, where $A$ is the amplitude determined by the collision velocity, $B$ is the damping coefficient describing the decay rate, and $\omega$ represents the vibration frequency. The values of $B$ and $\omega$ depend on the material properties, and we have chosen acrylic as the material with $B$ = 469 and $\omega/2\pi$ = 128 Hz. Such signals typically decay to imperceptible levels within 0.5 seconds, and we added additional control to stop the output after 0.1 seconds. The value of $A$ is determined by participants' perception, calibrated to be noticeable to most people while not being overwhelmingly intense. 

\item $F(t) = k \cdot n$ represents the pressure force, determined by the virtual coupling spring coefficient $k$ and the penetration depth $n$ of the fingertip into the rigid body. We selected $k = 100 \, \text{N/m}$ to provide adequate interaction stiffness for grasping while avoiding numerical instability and rapid force saturation associated with larger values. This value maps the device's perceptible force range (0.04 N to 1.36 N) to a practical penetration depth of 0.4 mm to 13.6 mm.

\item $S(t) = Q(t-t_1) + Q(t-t_2) + F(t)$ represents the vibration signal generated during the transition between static and kinetic friction in the stick-slip phenomenon, which is essentially an overlap of collision signals. $t_1$ is the moment of transition from static to kinetic friction \cite{konyo2008alternative}, while $t_2$ represents from kinetic to static. The parameters $F$, $v$, and the friction state are all calculated by the physics engine. In the case of dynamic friction, S(t) includes both the stick-slip transition signals and the sliding vibration feedback R(t). For clarity of presentation, Fig. \ref{feedbackimg} only shows the stick-slip signals.

\item $R(t) = F(t) \cdot v \sin(\omega t) + F(t)$ describes the vibration during sliding  \cite{adi2009haptic}. When sliding across a surface, microscopic surface roughness generates periodic vibrations. The intensity of these vibrations increases with both normal force $F$ and relative velocity $v$, as greater force or speed amplifies the perceived vibration. Therefore, we simplified the fingertip vibration perception during sliding as a signal that is proportional to both the normal force $F$ and relative sliding velocity $v$, with $\omega/2\pi$ representing the vibration frequency. In this study, we selected 100 Hz.
\end{itemize}

\subsection{Hardware evaluation}

\begin{figure}[!t]
  \centering
  \includegraphics[width=2.6in]{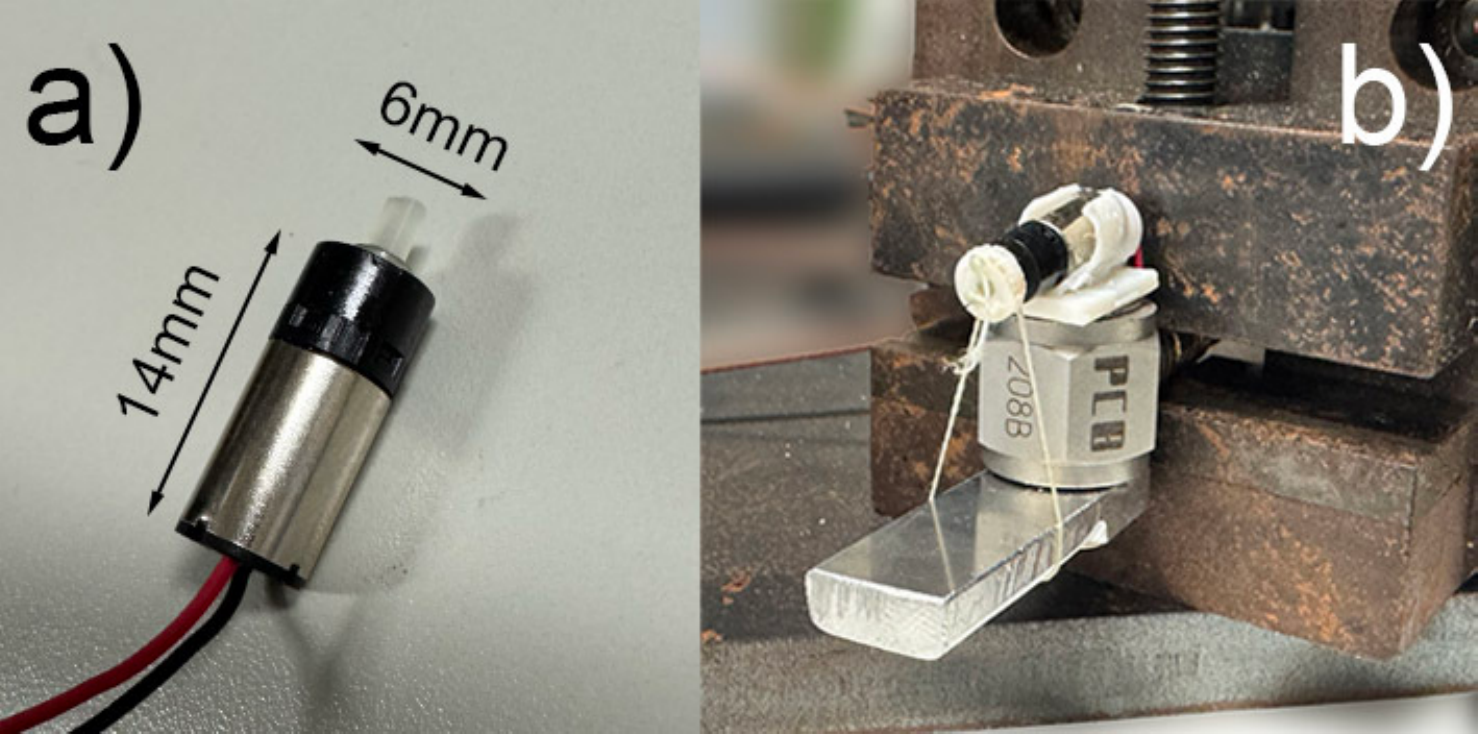}
  \caption{(a) The DC motor used in the device. (b) Experimental setup for measuring motor performance using a force sensor. This setup isolates the normal force transmitted by the string by using a rigid block and sensor, where the string pulls the sensor directly. It simulates the fingertip device configuration to measure force output and vibration characteristics. }
  \label{measureSet}
\end{figure}

We evaluated the device's hardware performance by measuring two key parameters: the motor's force output capability and its vibration frequency response. We aimed to measure how the motor's force output matches the software control and the resonance points for optimal strongest feedback. The measuring system is shown in Fig. \ref{measureSet}(b). A PCB Piezotronics 208C02 (PCB Piezotronics, Inc., USA) force sensor was used to measure the output force, with the measuring structure designed to simulate the fingertip device configuration. The system enabled the measurement of string pressure, saturation value, and frequency response through vibration intensity and current measurements. 

\begin{figure}[!t]
  \centering
  \includegraphics[width=2.2in]{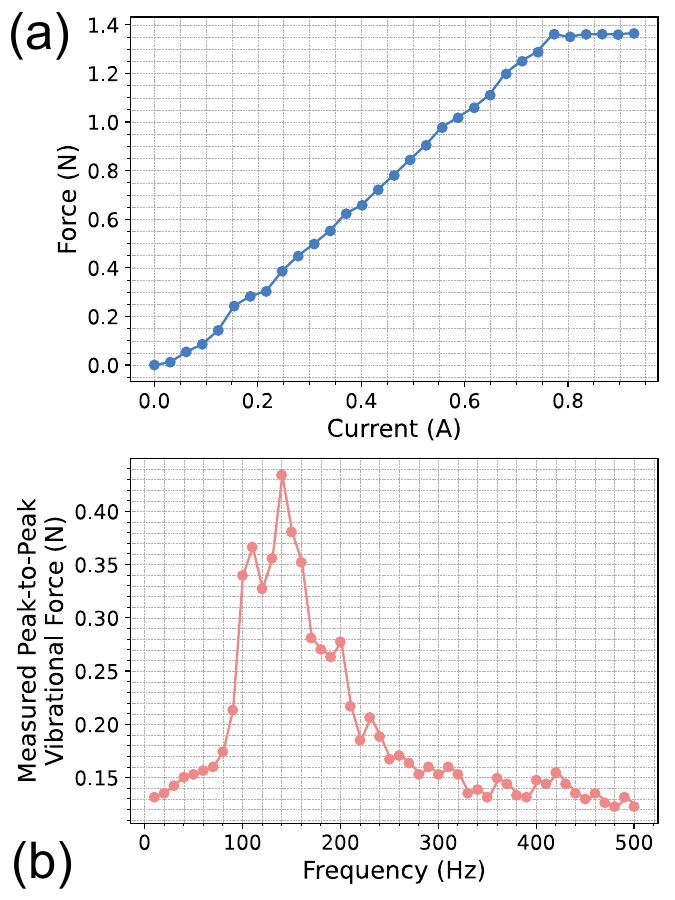}
  \caption{(a) Relationship between target current and measured force, showing how the motor's output force varies with the input current. (b) Frequency response characteristics of the motor-string system.}
  \label{MotorForce}
  \end{figure}

To assess the force output, we incrementally increased the motor's current output from 0 mA in steps of 30 mA, recording the force sensor values at each step. 
After each measurement, we checked the motor temperature, ensuring it had cooled down before proceeding to the next measurement.
The force measurement results are shown in Fig. \ref{MotorForce}(a). The pulling force for the motor reached a maximum of 1.36 N. The current and pulling force showed a generally linear relationship. According to \cite{aoki2009wearable}, the minimum perceivable fingertip pressure is approximately 40 mN. The proposed device delivers force significantly above this threshold while maintaining a slim design. The theoretical maximum force calculated from the stall torque (156.9 mN·m) and shaft radius (1 mm) is 1.57 N, the measured value is slightly lower, likely due to internal friction and the string's diameter increasing the effective winding radius.

To assess the frequency response of the vibration intensity, we input waveforms of a maximum of 500 Hz and measured the peak-to-peak value of the vibration intensity. A half peak-to-peak voltage DC bias was added to the signal to maintain unidirectional motor output, preventing mechanical noise caused by alternating directions of the motor. The peak-to-peak current of the signal was set to the value at which the motor reached its maximum force output. The frequency was increased in 10 Hz increments, with a cooling time between measurements. The results are shown in Fig. \ref{MotorForce}(b). The results indicate that the motor has a distinct resonance within the 100 Hz to 180 Hz range. The correlation coefficient of the linear fit between the input vibration frequency and the measured vibration frequency is 0.9998, with a mean squared error(MSE) of 3.70, indicating no significant deviation.

The device's weight is as follows: each fingertip unit weighs 1.55 g, while the hand-mounted components (circuit board and cables, excluding PC and power supply) weigh 11.29 g.

\section{Evaluation}

To assess our device's three key capabilities of providing essential haptic feedback for effective virtual manipulation while preserving natural finger dexterity in real-world interactions, we conducted a series of experiments after obtaining informed consent. These experiments were specifically designed to evaluate: 1) how well users perceive and respond to our haptic feedback elements (contact, grip force, and sliding vibration feedback) in virtual environments, 2) how these feedback mechanisms improve dexterous manipulation efficiency, and 3) whether our device design maintains unobstructed fingerpad interaction with real objects during daily tasks. 
A total of 12 participants (8 male and 4 female) took part in the study. The participants were aged between 22 and 48 (mean=27.92, SD=7.01). Nine participants reported prior VR experience. All participants chose to use their right hand as the preferred hand for the experiment. The experimental procedure received approval from the Institute of Science Tokyo Review Board (Approval Number: 2023362). 
During the virtual object interaction experiment, real-time data were collected. This included the positions of both the virtual object and the virtual rigid body hand, the force between the target object and the thumb fingertip, and the distance between the index and thumb fingertips.
Participants wore the proposed device on their preferred hand for single-hand tasks and on both hands for tasks requiring two-handed manipulation, and pose recognition was done using a Leap Motion Controller (UltraLeap Ltd., UK). For desktop scenes, the controller was placed on the table in front of the monitor's center, while for VR scenes, we mounted the controller on the headset. Participants were instructed to interact with virtual objects using the virtual rigid body hand. The physics engine's time step was 0.01 s (100 Hz), which represented a value that maintained system stability while keeping computational overhead relatively low. The haptic feedback system was capable of delivering feedback up to 500 Hz. For the display system, we used a monitor with a 180 Hz refresh rate for the fingertip pressure perception experiment and slip vibrotactile feedback experiment, while a Meta Quest Pro (Meta Platforms, Inc., USA) head-mounted display (HMD) with a 90 Hz refresh rate was used for the dexterous manipulation (peg-in-hole) experiment. Before the experiment, each participant had individual calibration. 
For pressure calibration, the current increased from 0 mA in 30 mA steps until the participant could just perceive it, setting this as the perception threshold. When the calculated output force is below this perception threshold (but greater than 0), the system outputs a force equal to the threshold to ensure users can always perceive feedback, even when the calculated value itself is below the perceptible level.
For contact vibration calibration, participants repeatedly tap virtual objects while freely adjusting the vibration amplitude until they find a feedback intensity that is both clearly perceptible and comfortable. During the questionnaire completion process, the experimenter faced away from the participants to encourage honest feedback. Participants were free to ask questions at any time during the filling process.


\subsection{Fingertip Pressure Perception Experiment}
\label{sec:presureExperiment}


Unlike many haptic devices that provide force feedback, our proposed device relies solely on string-delivered pressure feedback. We therefore investigated whether this continuous pressure feedback, without physical movement restrictions, could guide user perception and manipulation behavior. This approach differentiates our system from devices that offer only vibration feedback. It is crucial to assess user perception to determine whether they can truly feel the pressure and if this pressure can influence user behavior, resulting in reduced pinch distance.

\subsubsection{Methods}


\begin{figure}[!t]
\centering
\includegraphics[width=2.8in]{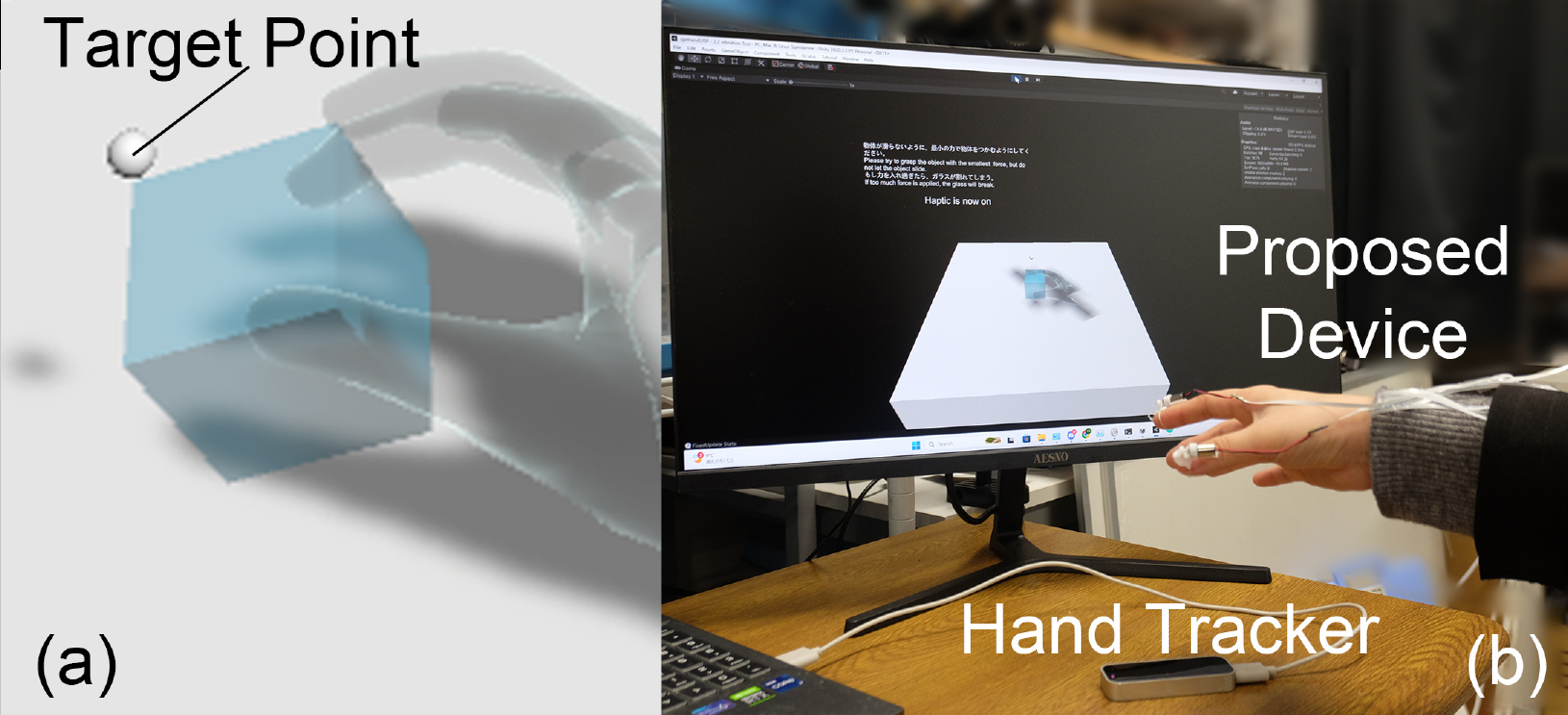}
\caption{The Fingertip Pressure Perception Experiment. (a) The Virtual Experiment Scene shows participants grasping a virtual glass cube and lifting it to a target height while applying minimal force to prevent dropping or breaking it. The cube provides visual feedback by changing color when properly positioned. (b) Overview of the experimental setup showing the physical arrangement of the haptic device, hand tracking, and display system. This experiment helps evaluate whether users can effectively perceive and respond to pressure feedback from the thin-string haptic device.}
\label{minForceExp}
\end{figure}


The setup is shown in Fig. \ref{minForceExp}. We aimed to test whether participants would apply different pinch distances when attempting to grasp objects with what they perceived as minimal force under different haptic feedback conditions.
During the experiment, participants wore the proposed device and were instructed to grasp the cube in the virtual environment.  They received a prompt: ``Please try to grasp the object with the smallest force, but do not let the object slide or drop. If too much force is applied, the glass will break." Participants were asked to grasp an object by thumb and index fingertips and determine the smallest grasp force that would prevent the ``glass" object from either falling due to insufficient force or breaking due to excessive force. The hand needs to be oriented with its dorsal side facing upward, with all five fingers fully extended, not flexed toward the palm. The index finger and thumb must remain clearly visible from a bottom-up perspective, without any occlusion to the hand tracker. Participants had to align the cube's center with the target point (height 10 cm) and maintain this alignment for at least three seconds to succeed. When the object's center was within 1.5 cm of the target point, the cube would gradually turn green, and after maintaining this position for 3 consecutive seconds, it would blink to indicate the completion of a single trial. We used a cube with a length of 5 cm and a weight of 300 g, assigning it a static friction coefficient of 0.3 and a dynamic friction coefficient of 0.1, values approximated based on glass properties. 
We set the cube to break in the physics engine when the total force applied by all fingers exceeded 8 N, a threshold chosen to ensure that participants needed to carefully control their grip without making the task excessively difficult. This breaking force was not directly fed back to the user's fingers but served as a constraint in the virtual environment. Also, the feedback tension depends on individual participants due to the calibration process.
During the experiment, the distance between the participant's thumb and index finger was continuously recorded.
Before the experiment, participants had time for training to become familiar with the environment. Participants needed to successfully complete 3 trials under each condition to finish the training phase. 
In the formal experiment, participants needed to complete 6 trials for each condition, and the order of trials was randomized. If the object was dropped or broken, it was recorded as a failed grasp, and the trial was not counted as complete and had to be repeated.
The experiment had three conditions: ``No Haptic", ``Contact Vibration", and ``Pressure". ``Contact Vibration" refers to the ``Collide" feedback shown in Fig. \ref{feedbackimg}, while ``Pressure feedback" refers to the ``Press" feedback shown in Fig. \ref{feedbackimg}.

\subsubsection{Results}
\begin{figure}[!t]
\centering
\includegraphics[width=3.5in]{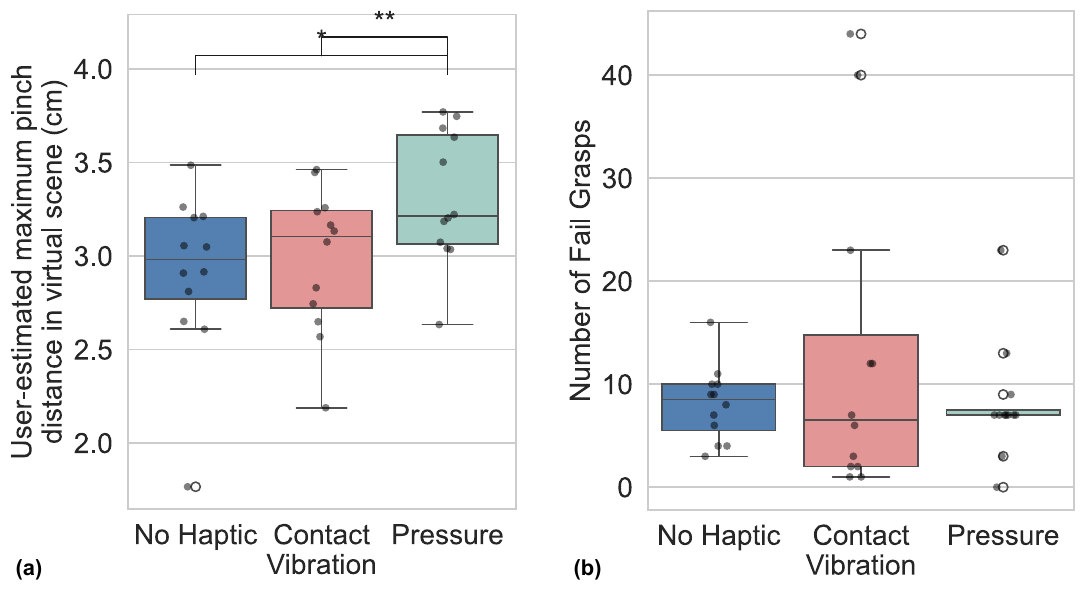}
\caption{(a) User-estimated minimum grip distance comparison across haptic feedback conditions. Results show pressure feedback led to larger grip distances compared to both No Haptic and Contact Vibration conditions, while no significant difference was found between No Haptic and Contact Vibration conditions. Under the presence of pressure, users maintained larger pinch distances without explicit instruction. (b) Analysis of failed grasps showed no significant differences in failure rates across conditions, with median failures of 8.50, 6.50, and 7.00 for No Haptic, Contact Vibration, and Pressure conditions, respectively (***p\textless 0.001, *p\textless 0.01, *p\textless 0.05).}
\label{minPressure1}
\end{figure}

The pinch distance under three different haptic feedback conditions is shown in Fig. \ref{minPressure1}(a). The pinch distance values were 2.91 cm (median=2.98, SD=0.442, SEM=0.128) for No Haptic condition, 2.98 cm (median=3.10, SD=0.387, SEM=0.112) for Contact Vibration condition, and 3.31 cm (median=3.21, SD=0.354, SEM=0.102) for Pressure condition. For each participant, we used their mean pinch distance across the 6 trials in each condition for statistical analysis. Statistical analysis revealed that all conditions showed normal distribution (Shapiro-Wilk test: No Haptic W=0.892, p\textgreater0.05; Contact Vibration W=0.944, p\textgreater0.05; Pressure W=0.928, p\textgreater0.05). A repeated measures ANOVA showed significant differences among conditions (F(2,22)=9.185, p\textless0.01). Post-hoc paired t-tests with Bonferroni correction revealed significant differences between Pressure and No Haptic conditions (p\textless0.05, t=-3.414) and between Pressure and Contact Vibration conditions (p\textless0.01, t=-4.034), while no significant difference was found between No Haptic and Contact Vibration conditions (p\textgreater0.05, t=-0.714).

The number of failed grasps under three different haptic feedback conditions is shown in Fig. \ref{minPressure1}(b). The number of failures were 8.08 (median=8.50, SD=3.630, SEM=1.048) for No Haptic condition, 12.75 (median=6.50, SD=15.094, SEM=4.357) for Contact Vibration condition, and 8.08 (median=7.00, SD=5.616, SEM=1.621) for Pressure condition. Statistical analysis revealed that while the No Haptic condition showed normal distribution (Shapiro-Wilk test: W=0.944, p\textgreater0.05), both Contact Vibration (W=0.751, p\textless0.01) and Pressure conditions (W=0.756, p\textless0.01) deviated from normality. Given these non-normal distributions, we employed the Friedman test, which revealed no significant differences among conditions ($\chi^2$(2)=0.553, p\textgreater0.05).

 \subsubsection{Discussion}

The results indicate that the proposed device's pressure feedback can change the behavior of the grasping task, suggesting that pressure feedback alone can enable users to find a relatively smaller grip distance. This outcome suggests that the proposed device can guide users to increase their pinch distance even without applying external forces to control finger movements. This improvement in performance with pressure feedback might be related to the continuous stimulation of SA1 receptors in the fingertips. When participants received this consistent pressure feedback, they likely gained a perception of being better able to maintain their grip on virtual objects, even though their finger movements were not physically constrained.
Moreover, applying a varying pressure gradient to the fingertips can simulate tactile changes during grasping. For example, gradually increasing pressure can represent the fingers gradually tightening their grip on the object.

On the other hand, there was no significant difference between ``Contact Vibration" and ``No Haptic" feedback. The condition ``Contact Vibration" can provide information about the start of contact, but does not convey continuous information about the applied force. ``Contact Vibration" alone may not be sufficient to alter users' behavior in grasping tasks. The limited effectiveness of contact vibration may also be attributed to its short duration. Contact vibration typically occurs only at the initial moment of contact, providing a brief, instantaneous signal. This feedback may not allow sufficient time for participants to process the tactile information and integrate it into their grasping behavior.

While there was no statistically significant difference in the number of failed grasps, it is noteworthy that in the Pressure condition, failures tended to cluster around 7 attempts. This number closely aligns with the required 6 trials per condition. This pattern suggests that pressure feedback may have been effective. Participants typically used about one failed attempt per trial to more precisely calibrate their perception of the applied pressure.

\subsection{Slip Vibrotactile Feedback Experiment}
Sliding feedback during object grasping is one of the important properties of dexterous manipulation. In our previous research \cite{xu2024optimizing}, we observed that participants could grasp an object with their fingers, gently release it to allow sliding, and then immediately grasp it again using only vibration feedback, even in the absence of tangential force. This work also noted that there was no increase in reaction delay compared to situations where tangential force is presented. The device proposed in the current paper cannot display tangential force. Therefore, we investigated whether vibration or pressure alone, generated by the proposed device, can significantly improve users' reaction time when an object slides. We added pressure feedback experiments to test whether pressure feedback could also indicate the sliding reaction time to users.

\subsubsection{Methods}

\begin{figure}[!t]
\centering
\includegraphics[width=3.2in]{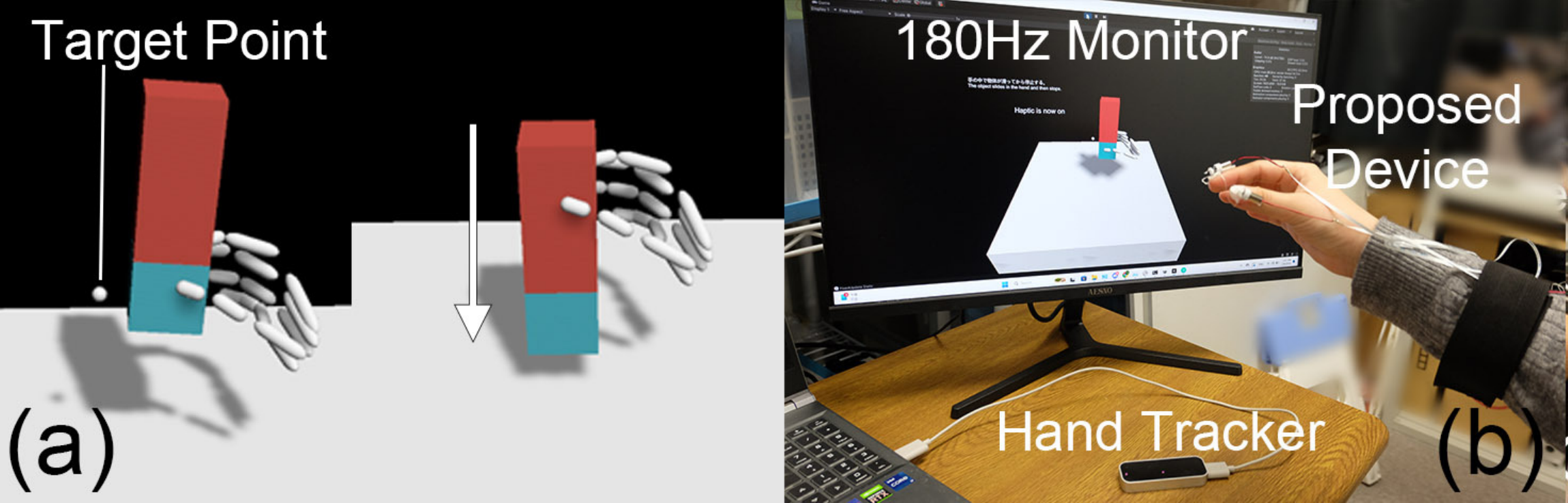}
\caption{Experimental setup for evaluating slip detection and response. (a) Virtual environment showing the task where participants grasp, slide, and attempt to catch a falling object. The blue area indicates the grasping region, where participants were instructed to grasp the object's blue part with their fingertips and align the center of the blue part with the target point before proceeding. (b) Hardware setup demonstrating the high-refresh-rate display system used to provide precise visual feedback for measuring participants' reaction times to object sliding.}
\label{slidingExp}
\end{figure}

The setup is shown in Fig. \ref{slidingExp}. Participants were asked to slide an object and re-grasp it before it fell to the ground. We used a virtual cuboid of 15 cm in length, 5 cm in width, and 5 cm in height. The static friction coefficient was set to 0.15, the weight was 300 g, and the dynamic friction coefficient was set to 0.10 for easy sliding. One-third of the object's length at one end was colored blue, with its center of mass aligned with the blue section's center to minimize rotation during sliding and facilitate re-grasping. A high refresh rate monitor (180 Hz) was used to provide more accurate visual feedback of the object's movement, particularly at the moment of slipping, which was crucial for measuring reaction time.


Participants were instructed to grasp the blue area and align it with the target point before the slide started. Participants were asked to grasp with their thumb and index fingertips, with all five fingers fully extended, not flexed toward the palm. The index finger and thumb must remain clearly visible from a bottom-up perspective, without any occlusion to the hand tracker. When the center of the blue part was within 1.5 cm of the target point, the blue color would fade, indicating that the participant could begin releasing. The trial began when the user felt they had a stable grasp on the object and were ready to start releasing it. A successful grasp was recorded if participants could stabilize the object before it hit the ground. To ensure valid trials, participants were prohibited from quickly releasing and immediately pinching their fingers without feedback. Invalid attempts were identified by two characteristics: 1) rapid finger opening and closing at a similar speed without monitoring the sliding moment, which could prevent the object from falling but indicated a misunderstanding of the task rules, and 2) zero contact force between fingertips and object during the fall, indicating the fingers were released too much to receive any haptic feedback. Such attempts were immediately flagged, and participants were asked to retry. Trials in which participants failed to respond to the sliding object due to occasional distraction were also considered invalid.

Before the formal experiment, participants underwent pre-training for each condition. For each condition, participants were required to demonstrate their understanding of the rules and complete three consecutive valid trials before proceeding. Once participants had successfully completed pre-training across all conditions and indicated they were ready to begin, the formal experiment commenced. The experiment had 4 conditions:``No Haptic", ``Pressure", ``Pressure and Vibration", and ``Vibration". Pressure feedback corresponded to the ``Press" shown in Fig. \ref{feedbackimg}, while vibration feedback included both ``Slide" and ``Stick-slip" shown in Fig. \ref{feedbackimg}. Each condition consisted of 8 trials, and the order of trials was randomized. We measured response latency as the time difference between when the object started sliding (T1) and when the fingers exhibited a grasping response (T2). We recorded T2-T1 for each attempt and considered it the response latency.

\begin{figure}[!t]
\centering
\includegraphics[width=3.5in]{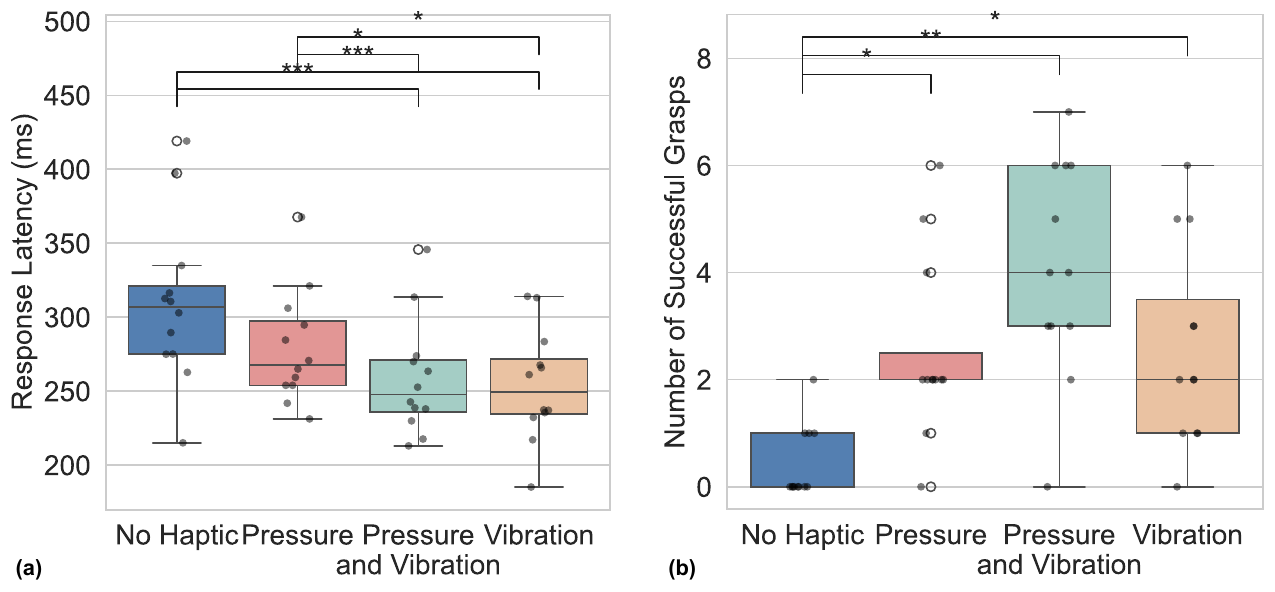}
\caption{(a) Comparison of the response times to detect object sliding across four haptic feedback modes. There are significantly shorter response times when vibration is included, compared to relying on pressure alone or having no haptic feedback. However, pressure feedback did achieve faster reaction times than no haptic feedback in some participants. (b) A comparison of the number of successful grasps under the same four conditions shows that both pressure and vibration feedback stand out by helping users rapidly re-grasp the object, while the absence of haptic cues leads to the lowest success rate.
}
\label{slidingSuccessRate}
\end{figure}

\subsubsection{Results}

Sliding detection response times under different haptic feedback conditions are shown in Fig. \ref{slidingSuccessRate}(a). For each participant, we calculated their mean response time across the 8 trials in each condition. The response times were 309.29 ms (median=306.80 ms, SD=55.87, SEM=16.13) for No Haptic condition, 279.19 ms (median=267.84 ms, SD=38.40, SEM=11.08) for Pressure condition, 258.30 ms (median=247.74 ms, SD=38.94, SEM=11.24) for combined Pressure and Vibration condition, and 254.17 ms (median=249.27 ms, SD=37.93, SEM=10.95) for Vibration condition. Statistical analysis confirmed normal distribution across all conditions (Shapiro-Wilk: all p\textgreater0.05). A repeated measures ANOVA revealed significant differences among conditions (F(3,33)=22.75, p\textless0.001). Post-hoc paired t-tests with Bonferroni correction showed that both vibration-only and pressure-and-vibration conditions significantly improved response times compared to no haptic condition (p\textless0.001 for both). Additionally, both vibration-only and pressure-and-vibration conditions showed significantly better performance than pressure-only conditions (p\textless0.05 for both).

The number of successful grasps of each participant also varied across conditions are shown in Fig. \ref{slidingSuccessRate}(b), with means of 0.42 (median=0.00, SD=0.67, SEM=0.19) for No Haptic, 2.50 (median=2.00, SD=1.68, SEM=0.48) for Pressure, 4.08 (median=4.00, SD=2.02, SEM=0.58) for Pressure and Vibration, and 2.58 (median=2.00, SD=1.88, SEM=0.54) for Vibration condition. Due to non-normal distributions in some conditions (Shapiro-Wilk: p\textless0.05), we conducted a Friedman test, which revealed significant differences among conditions ($\chi^2$=21.50, p\textless0.001). Post-hoc Wilcoxon signed-rank tests with Bonferroni correction showed that all haptic feedback conditions significantly improved success rates compared to the No Haptic condition (all p\textless0.05).

\subsubsection{Discussion}
The results showed that vibration could significantly reduce participants' response time when re-grasping a sliding object. This conclusion aligns with our previous research \cite{xu2024optimizing}, demonstrating that the proposed device can improve dexterous manipulation efficiency through sliding vibration feedback. However, pressure feedback alone is insignificant in improving users' reaction time in this specific task. This may be because the current pressure feedback only displays normal force and does not change with object sliding. During the experiment, participants often say, ``The object feels completely smooth, and it is difficult to perceive the sliding" when switching to the pressure-only mode. If changes in tangential force during sliding could also be presented by pressure, which could be adjusted through changes in haptic rendering design, it might also contribute to this task. However, the current result is noteworthy in that many participants showed better response times under the Pressure condition compared to the No Haptic condition. This finding indicates that the current device's pressure intensity can at least partially indicate when the object is about to slip due to insufficient grip force, although not as effectively as vibration. 

Compared to our previous study \cite{xu2024optimizing}, the current study shows longer reaction times, likely due to increased latency from the hand tracker and software. The overall higher latency can be attributed to two main factors: 1) The hand tracker data processing and complete physical hand simulation in the physics engine require significant computational time, with measured delays of approximately 70 ms, and 2) Different graphics and physics settings result in less obvious visual and haptic changes during the initial sliding phase, requiring a longer time before being perceived. The reaction time difference between No Haptic and Vibration conditions was around 90 ms in the previous study, but only around 60 ms in the current one. We speculate this could be due to two reasons: 1) The previous study required successful grasping, collecting only the fastest reaction data, while the current experiment only observed the pinch reflex without requiring successful manipulation. 2) The small device's weak vibrations may be difficult to perceive during the low-speed initial sliding phase, increasing reaction delays. Additionally, previous studies involved grasping real objects (the handle of the device), while the current experiment only requires finger movement to decrease distance, potentially causing differences.

However, regarding the number of successful grasps, we found that conditions with pressure feedback improved success rates. This result suggests that continuous pressure feedback, by activating SA1 receptors in the skin, enhances users' perception of the grasping state and enables real-time force adjustments during object manipulation. This phenomenon also reflects that while sliding vibration feedback is effective for indicating the moment of initial contact, vibration signals alone may not provide sufficient information. Pressure feedback may compensate for this limitation.

\subsection{Evaluation of Dexterous Manipulation Task}

We conducted experiments to evaluate both the device's impact on dexterous manipulation efficiency and users' subjective experiences. We designed an evaluation to gain insights into participants' thoughts and experiences. First, we recorded participants' operation times during virtual object manipulation tasks. Second, participants evaluated their perception of different haptic feedback components, including Pressure, Slide, and Contact, as well as their overall sense of acceptability of the haptic feedback. Third, we assessed the device-wearing experience in terms of comfort and convenience, considering factors like weight, size, wearing process, and potential fatigue or discomfort.

\subsubsection{Methods}

\begin{figure}[!t]
\centering
\includegraphics[width=2.5in]{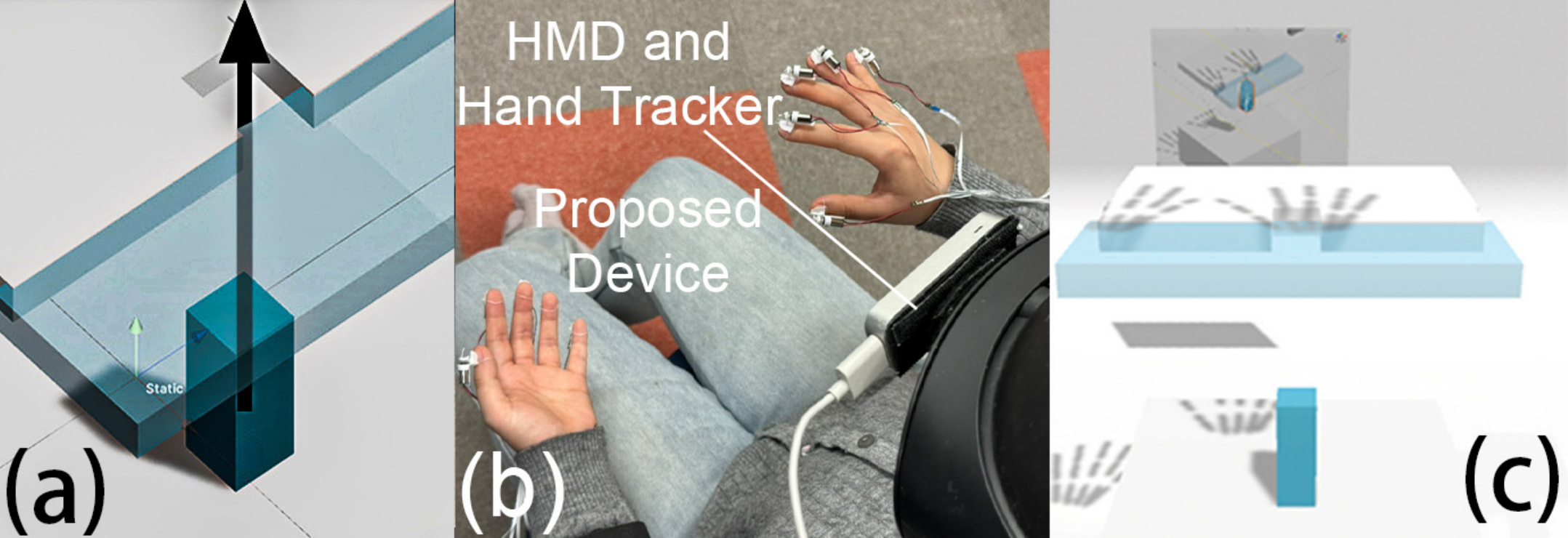}
\caption{The peg-in-hole task: a dexterous manipulation benchmark. (a) The experiment virtual scene shows a cuboid object placed on a platform that needs to be inserted through a hole in a higher platform. The hole is slightly larger than the object's cross-section. Part of the platform is made semi-transparent to help participants better recognize the insertion process. (b) The experimental setup utilizes an HMD for improved depth perception during manipulation, along with a Leap Motion Controller installed on the HMD for hand tracking. (c) A demonstration video plays in the virtual environment, showing only the object's movement path, without visible hands, to show the task and avoid biasing participants' manipulation strategies.
}
\label{pegInHoleExp}
\end{figure}

We conducted a variation of the peg-in-hole task, as shown in Fig. \ref{pegInHoleExp}. The platform was 80 cm above the ground. The object to be manipulated measured 3.5 cm in length, 3.5 cm in width, and 10 cm in height. The setup included a second platform with its lower surface positioned 24 cm above the first platform. The second platform had a thickness of 2.5 cm and contained a through square hole. The hole on the platform has a square shape with dimensions of \(3.95 \, \text{cm} \times 3.95 \, \text{cm}\).
 A part of the platform is semi-transparent, allowing participants to see the inner surface of the hole. This design was made to evaluate the efficiency of the proposed lightweight haptic device for dexterous manipulation under visibility conditions. We used an HMD since it provides stereo vision. We utilized the Meta Quest Pro as the display hardware and still attached a Leap Motion Controller, as it performed better than the built-in hand tracking. In this experiment, participants wore the proposed device on both hands to enable two-handed manipulation. In the virtual environment, there is an area where a demonstration video of the peg-in-hole operation is repeated. The video only shows the translation of the cuboid, and no hands are displayed to avoid hinting at behavioral cues.

Users had practice time and were prompted with the phrase, ``Find a comfortable strategy to complete this task". The experiment had two conditions: haptic on and off. The ``on" condition enabled all feedback types shown in Fig. \ref{feedbackimg}, while ``off" disabled all feedback. The training ended only after users completed 3 trials under each condition. 
Then the formal experiment began with 8 trials per condition, and the order of trials was randomized. If the object fell out of the scene, it would be reset to its initial position and counted as a failure. If the object was accidentally dropped during grasping, it was also considered a failure. The trial was considered successful when the object completely exited the hole without falling back. We allowed users to intentionally place the object down to adjust their grasp points. This was not counted as a failure, but the timing would not be reset. Timing began when the participant first contacted the object and ended when the object completely exited the hole. The manipulation time of each trial was recorded.

After the peg-in-hole experiment was completed, participants were given an anonymous questionnaire (Table \ref{tab:questionnaire}). Participants rated these statements, presented in random order, using a seven-point Likert scale (1 = completely disagree, 7 = completely agree). 

\begin{figure}[!t]
\centering
\includegraphics[width=3.5in]{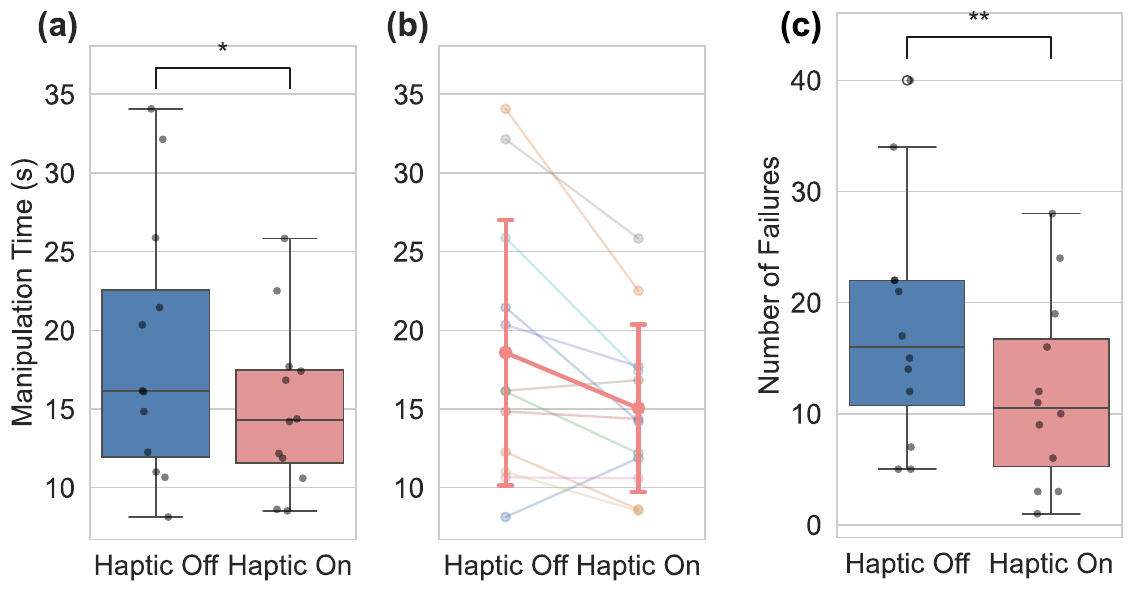}
\caption{Peg-in-hole task performance comparison between haptic feedback conditions. (a) Box plot of the manipulation time data for all participants. Results demonstrated that haptic feedback slightly improved task performance, with notably reduced variability. (b) Individual Change Plot showing each participant's performance across trials. (c) A box plot comparing the number of failed attempts, showing that haptic feedback substantially reduced task failures across participants.}
\label{peg}
\end{figure}

\subsubsection{Results}

The peg-in-hole task completion time results are shown in Fig. \ref{peg}(a)(b). For each participant, we calculated their mean completion time across the 8 trials in each condition. Task completion times for the haptic feedback condition (median=14.29 s, SD=5.32 s, SEM=1.54 s) were lower than those without haptic feedback (median=16.12 s, SD=8.44 s, SEM=2.44 s). Given the normal distribution of differences between these mean completion times across conditions (Shapiro-Wilk: p\textgreater0.05), a paired t-test revealed a statistically significant improvement in performance with haptic feedback (t=2.84, p\textless0.05).

The number of failed attempts during the task is shown in Fig. \ref{peg}(c). With haptic feedback, participants had fewer failures (median=10.50, SD=8.52, SEM=2.46) compared to those without haptic feedback (median=16.00, SD=10.88, SEM=3.14). A paired t-test showed significantly fewer failures with haptic feedback (t=4.28, p\textless0.01).

As shown in Table \ref{tab:questionnaire}, participants reported high scores for perceiving all three types of haptic feedback provided by the device (Q1-Q3, median\textgreater6, IQR\textless1). Moderate scores were given for overall acceptability (Q4, median=4.5, IQR=2.25). They strongly agreed that the haptic feedback aided task completion (Q5, median=6, IQR=0.25). While participants found the peg-in-hole task moderately challenging (Q6, median=5, IQR=1). Regarding wearability, participants rated the device's weight and size as highly suitable for daily wear (Q7-Q8, median=7, IQR=1) and noted quick response times without noticeable latency (Q9, median=6.5, IQR=1). Participants reported moderate fatigue levels during extended use (Q10, median=4, IQR=3.25), suggesting potential challenges for prolonged wear despite the absence of physical support during gesture-based interactions. However, moderate scores were given for wearing convenience (Q11, median=2.5, IQR=1). Participants rated the device's comfort level as moderate (Q12, median=3, IQR=2).

\begin{table}[!t]
  \caption{Participant Experience Evaluation of the Haptic Device\label{tab:questionnaire}}
  \centering
  \begin{tabular}{|p{6.2cm}|c|c|}
  \hline
  Question & Median & IQR\\
  \hline
  Q1: When using this device, I can clearly perceive the contact force. & 6 & 1\\
  \hline
  Q2: When using this device, I can clearly perceive the contact vibration. & 7 & 1\\
  \hline
  Q3: When using this device, I can clearly perceive the sliding vibration. & 7 & 0\\
  \hline
  Q4: The overall sense of acceptability when manipulating virtual objects with this device. & 4.5 & 2.25\\
  \hline
  Q5: Haptic feedback helps with manipulation in the Peg-in-hole task. & 6 & 0.25\\
  \hline
  Q6: The overall difficulty of the peg-in-hole task is moderate. & 5 & 1\\
  \hline
  Q7: The weight of the device is suitable for daily wear. & 7 & 1\\
  \hline
  Q8: The size of the device is suitable for daily wear. & 7 & 1\\
  \hline
  Q9: The device responds quickly. & 6.5 & 1\\
  \hline
  Q10: The level of fatigue when using the device for an extended period. & 4 & 3.25\\
  \hline
  Q11: This device is convenient to wear. & 2.5 & 1\\
  \hline
  Q12: The overall comfort level when wearing this device. & 3 & 2\\
  \hline
  \end{tabular}
\end{table}

\subsubsection{Discussion}
In the peg-in-hole experiment, we identified several key factors. First, some users mentioned that ``the lack of haptic feedback increases the difficulty of the operation" during the experiment.
According to the result of the fingertip pressure perception experiment, haptic feedback may help users avoid applying excessive or unbalanced forces when grasping objects, thereby reducing the probability of grasping failure.
Second, during the process, when users attempt to gently insert a misaligned peg into the hole, although there is no vertical tension feedback, the vibration cue indicates that the peg has collided with the outside of the hole, prompting users to adjust the position. 
Furthermore, once the peg enters the hole, tension feedback assists users in precisely guiding the cuboid. 
The significant reduction in the number of failures with haptic feedback is likely due to these factors. We also noticed that users often use one hand to insert the object into the hole while gently pushing the object with the other hand. Haptic feedback allows users to accurately apply force to the object with their other hand without needing to shift their visual focus away from the hole, reducing operation time. 

However, the accuracy of optical hand recognition (Leap Motion Controller) is insufficient for precise operations, with occlusion issues inherent in camera-based tracking causing frequent failures when fingers occluded each other or when the palm blocked visibility. These tracking issues increased randomness in the results. Nevertheless, the experimental data indicate that the haptic feedback generated by our device improves the efficiency of dexterous manipulation. However, in the absence of force feedback, haptic feedback has a limited effect on adjusting the object to align with the hole. Applying tangential forces with the fingers to determine the contact between the peg and the hole is still crucial.

Based on the subjective ratings, participants reported high scores for perceiving all three types of haptic feedback provided by the device. While participants found the peg-in-hole task moderately challenging, likely due to imperfect haptic feedback compared to reality and hand-tracking inaccuracies, they strongly agreed that the haptic feedback aided task completion. Regarding wearability, participants rated the device's weight and size as highly suitable for daily wear and noted quick response times without noticeable latency. However, moderate scores were given for several aspects. First, participants rated the overall feedback acceptability as moderate, which may be attributed to the device's inability to provide tangential forces and multi-point/edge feedback, reducing the realism of interactions. Second, the wearing convenience is possibly due to the challenge of managing multiple fingertip devices that can become tangled and require individual finger identification during wear. Additionally, although fatigue levels were relatively low, the lack of physical support during prolonged gesture-based interactions in mid-air may have contributed to these moderate ratings.

\subsection{Real-World Task Experiment and Subjective Experience Evaluation}
To evaluate the device's ability to preserve natural tactile perception and finger dexterity during real-world interactions, we designed experiments comparing user performance across different device-wearing conditions. While various haptic feedback approaches exist in the literature, we focused on comparing them with a thin glove condition as a representative case of devices that can provide fingertip pressure but cover the palmar side of fingers, affecting tactile perception. We selected two representative tasks requiring precise finger movements and surface detail perception: nut tightening and typing. These tasks require users to rely on natural tactile perception of surface details, such as nut surface features and keyboard key edges. We recorded completion times and collected subjective evaluations across different conditions.

\subsubsection{Methods}

\begin{figure}[!t]
  \centering
  \includegraphics[width=3.2in]{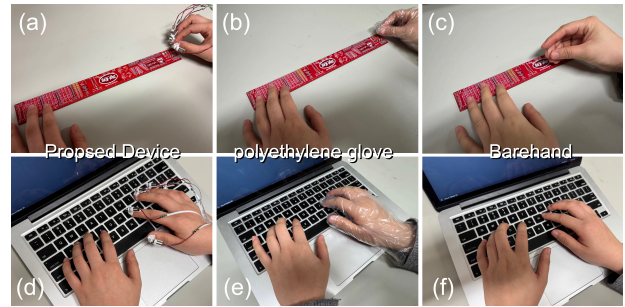}
  \caption{Real-world interaction tasks under three conditions: wearing the proposed haptic device (a,d), wearing a thin polyethylene glove (b,e), and barehand (c,f). The experiment consisted of two tasks: a nut-tightening task (a-c) requiring precise manipulation of small components, and a typing task (d-f) involving keyboard interaction. These tasks were designed to evaluate how different hand conditions affect real-world object manipulation.}
  \label{3types}
  \end{figure}

The three conditions tested were barehand, wearing the proposed device, shown in Fig. \ref{3types}, and wearing a polyethylene glove with a thickness of 0.023 mm, which fit closely to the fingertips when wearing. For the nut-tightening task, standardized M-2 screws (2 mm inner diameter, 20 mm in length) and M-2 nuts (1.5 mm in thickness) were prepared. In the typing task, participants sat at a desk using a MacBook Pro A1502 laptop (Apple Inc., USA) with English input and were required to type three different pangrams (a sentence containing every letter of the alphabet at least once), each consisting of exactly 33 letters. Before the formal experiment, participants underwent a training phase to familiarize themselves with both tasks under all three conditions. For the nut-tightening task, participants practiced three times under each condition before the formal experiment. For the typing task, participants practiced once under each condition using a different pangram text other than those used in the formal experiment. In the formal experiment, under each wearing condition, participants first completed the nut-tightening task, followed by the typing task. Each participant completed one trial per condition, and the order of trials was randomized. For the nut-tightening task, the screw was fixed on a 30 cm ruler with holes, and the nut was placed in the center of the ruler. The screw was placed on the same side as the participant's preferred hand, and the non-operating hand was only allowed to press the ruler and could not touch the screw or nut.  Timing starts when the first contact with the nut is made and stops when the nut is fully tightened onto the screw and the hand is removed. For the typing task, the glove was worn on the preferred hand, the three pangram texts were randomly assigned to the three conditions for each participant. Timing started with the entry of the first character and stopped upon completion of the last character.

After this experiment, participants were asked to fill out a questionnaire. The main question was ``Regarding the experience of operating in the real world while wearing a haptic device, please score each experience." Participants were to give a score for each condition on a scale from 1 (very poor) to 7 (excellent).

\subsubsection{Results}

\begin{figure}[!t]
\centering
\includegraphics[width=3.5in]{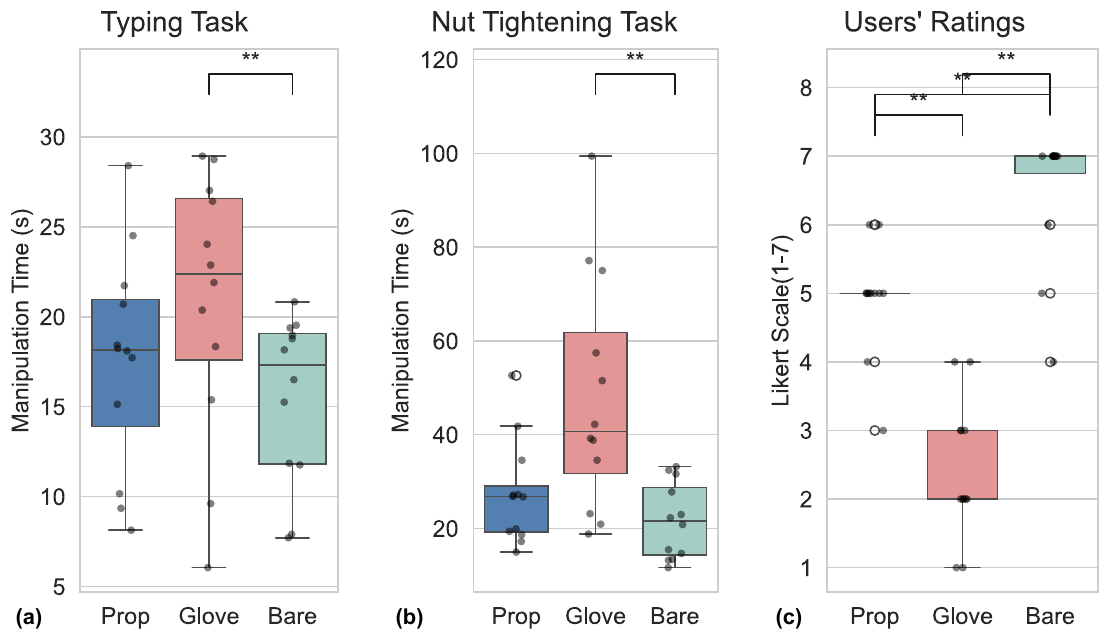}
\caption{Real-world interaction task performance comparison across three conditions: barehand (Bare), proposed device (Prop), and thin polyethylene glove (Glove). (a) Participants showed comparable typing performance across all conditions. (b) In the nut-tightening task, while the glove condition showed notably longer completion times, the proposed device maintained performance levels similar to barehand operation. (c) Subjective ratings demonstrated clear user preference for barehand over both the proposed device and glove conditions, with the proposed device receiving significantly higher ratings than the glove condition.}
\label{manipulationTotal}
\end{figure}

Manipulation task results are shown in Fig. \ref{manipulationTotal}. In the typing task (Fig. \ref{manipulationTotal}(a)), mean completion times were 17.55 s (median=18.17, SD=6.12) for the proposed device, 20.81 s (median=22.39, SD=7.35) for the glove, and 15.55 s (median=17.33, SD=4.64) for the barehand condition. The Shapiro-Wilk test showed normal distribution for all conditions (all p\textgreater0.05). A repeated measures ANOVA revealed significant differences among conditions (F(2,22)=8.29, p\textless0.01). Post-hoc paired t-tests with Bonferroni correction showed significant differences between glove and barehand conditions (p\textless0.01), while other comparisons were not significant.

For the nut-tightening task (Fig. \ref{manipulationTotal}(b)), mean completion times were 27.24 s (median=26.76, SD=11.08) for the proposed device, 48.17 s (median=40.69, SD=25.05) for the glove, and 21.62 s (median=21.55, SD=8.07) for barehand. The Shapiro-Wilk test showed normal distribution for all conditions (all p\textgreater0.05). A repeated measures ANOVA revealed significant differences among conditions (F(2,22)=9.54, p\textless0.001). Post-hoc paired t-tests with Bonferroni correction showed significant differences between glove and barehand conditions (p\textless0.01), while other comparisons were not significant.

The subjective evaluation (Fig. \ref{manipulationTotal}(c)) revealed significant differences among conditions (Friedman test: $\chi^2=23.00$, p\textless0.001). Post-hoc Wilcoxon signed-rank tests with Bonferroni correction showed significant differences between all pairs of conditions (all p\textless0.01). Mean ratings were 4.92 (median=5.00, IQR=0) for the proposed device, 2.42 (median=2.00, IQR=1.00) for the glove, and 6.50 (median=7.00, IQR=0.25) for barehand, indicating participants strongly preferred barehand and the proposed device over the glove condition.

\subsubsection{Discussion}
In the typing tasks, statistical tests found no significant differences between the barehand condition and the proposed device. This indicates that the device preserves much of the natural tactile feedback necessary for efficient keyboard interaction. While the barehand condition achieved the fastest completion time (17.33 s), the proposed device (18.17 s) performed a bit better than the glove condition (22.40 s). In contrast, the decline in performance observed in the glove condition suggests that thin plastic gloves can substantially impair typing efficiency by reducing tactile sensitivity to key edges and positions. Although we tried to fit the gloves closely to the fingertips, we could not provide perfectly fitted gloves customized for each participant's finger size. Additionally, even with close fitting, air pockets inside the gloves and the deformable nature of thin gloves likely affected users' ability to accurately detect key edges during typing. 

In the nut-tightening task, the completion time while wearing the glove (40.69 s) was more than twice that of barehand operation (21.55 s), with statistically significant differences between the glove and both the proposed device and barehand conditions. This suggests that gloves severely impede manipulation, making precise operations significantly more challenging. The performance of the proposed device (26.76 s) was not significantly different from the barehand condition, indicating that the device preserves crucial tactile feedback and finger dexterity required for manipulating small objects.

In the subjective rating, statistically significant differences were found between all conditions. The proposed device (5.00) received notably higher ratings than the glove condition (2.00), suggesting that users perceived clear advantages in tactile sensitivity and operational comfort. While participants still preferred barehand operation (7.00), the relatively high rating of the proposed device indicates that it provides an acceptable balance between barehand and glove conditions.

The nut-tightening results indicate that, compared to glove-type devices, the proposed device allows users to complete precise manipulation tasks with less hindrance. The performance similarities to barehand operation suggest that users can maintain interaction capabilities while wearing the device, compared to glove-type devices. 
Furthermore, while gloves represent only a subset of haptic devices, our comparison mainly focused on devices that generate pressure feedback on the fingerpad. Additional comparisons with other types of haptic devices, such as fingertip-worn devices, would be valuable for evaluation in the future.

\section{Limitations} 
Our study revealed several limitations. 
One limitation is the small contact area of the string, which differs from the distributed pressure experienced when grasping real objects. However, research has shown that localized haptic stimuli, such as pin arrays or even substitutive vibrotactile feedback \cite{ooka2010virtual}, can effectively convey complex interaction cues like force and slip. Our experiments confirm that the feedback provided by the string, while localized, is functionally sufficient to enhance manipulation performance.
Furthermore, the design still restricts finger movement during operation by cables, and while the device is more compact than traditional haptic devices, the string mechanism still imposes more physical constraints than fingerpad-free approaches \cite{preechayasomboon2021haplets}. Several engineering challenges persist. For instance, during interaction, accidental contact with external real objects could exert mechanical stress on the string-winding structures. Additionally, the device requires users to clean their fingernails before installation.
The current infrared stereo camera-based tracking system shows insufficient accuracy for precise operations due to vision occlusion. Also, our participant pool lacked diversity in VR experience, age, and gender distribution. The limited number of trials in the real-world task experiment may affect the reliability of the results. 
Additionally, the time-consuming device mounting process affected user experience, suggesting the need for a design to reduce the long process of identifying and mounting individual devices for each finger. Finally, some users reported heavy reliance on visual feedback during dexterous manipulation tasks, indicating room for improving haptic feedback, particularly in multi-directional information.

\section{Conclusions}
In this study, we developed a lightweight, wearable haptic device that provides physics-based haptic feedback for dexterous manipulation in virtual environments without hindering real-world interactions. By employing thin strings and attaching actuators to the fingernails, the proposed device delivers tension-based haptic feedback at the fingertips while preserving the ability to interact with physical objects.
Through a series of experiments, we investigated the effectiveness of the proposed device in various scenarios. The fingertip pressure perception experiment demonstrated that participants could perceive and respond to the pressure feedback, the pinch distance can be larger than no haptic feedback when a virtual object is in hand. The slip vibrotactile feedback experiment showed that sliding vibration feedback improved their reaction time to the sliding task. The feedback can also significantly improve the efficiency of dexterous manipulation in the peg-in-hole experiment.
The real-world task experiments showed its ability to preserve natural tactile sensations and small hindrances to real-world operations. By keeping the palmar side of the fingers free, the device allows users to perform common tasks like typing and nut-tightening with similar performance to barehanded interactions.
The subjective evaluation provided insights into participants' experiences. Most participants agreed with the enhanced realism and precision provided by the haptic feedback, especially in judging the gripping state and adjusting the applied force. 
However, moderate scores were given for overall acceptability and wearing convenience, suggesting room for improvement in the device's design. Despite these limitations, participants reported relatively low fatigue levels during extended use and rated the device's weight and size as highly suitable for daily wear. Overall, our results demonstrate that the proposed fingertip-based haptic device successfully enhances virtual manipulation experiences while preserving natural tactile feedback for real-world interactions.

\bibliographystyle{ieeetr}
\bibliography{ref-base}

\vspace{-35pt}

\begin{IEEEbiography}[{\includegraphics[width=1in,height=1.25in,clip,keepaspectratio]{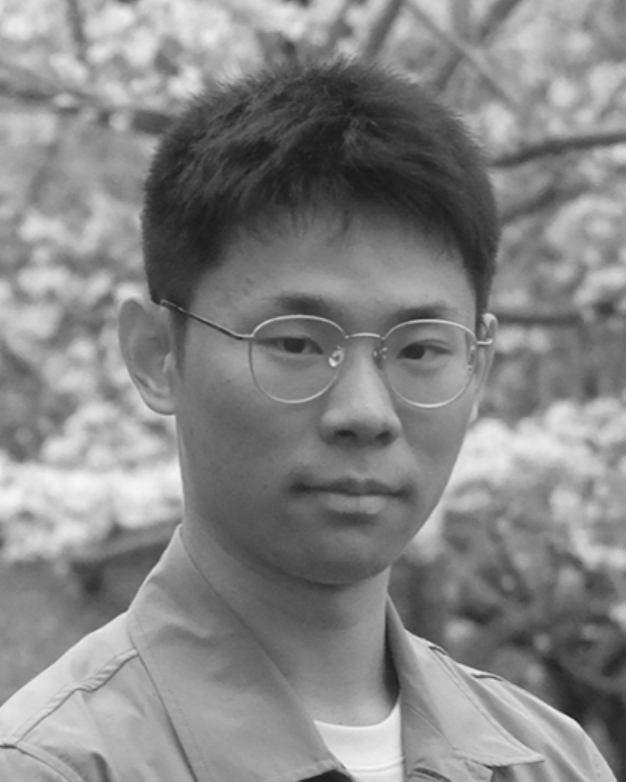}}]{Yunxiu XU} received the B.S. degree from Beijing Information Science and Technology University, Beijing, China, and the M.S. degree in Engineering from the Department of Information and Communication Engineering, Institute of Science Tokyo, Yokohama, Japan, in 2022. He is currently pursuing the Ph.D. degree at the same institution. He is a Research Fellow (DC2) of the Japan Society for the Promotion of Science (JSPS). His research focuses on haptics and interaction in virtual reality.
\end{IEEEbiography}

\vspace{-32pt}

\begin{IEEEbiography}[{\includegraphics[width=1in,height=1.25in,clip,keepaspectratio]{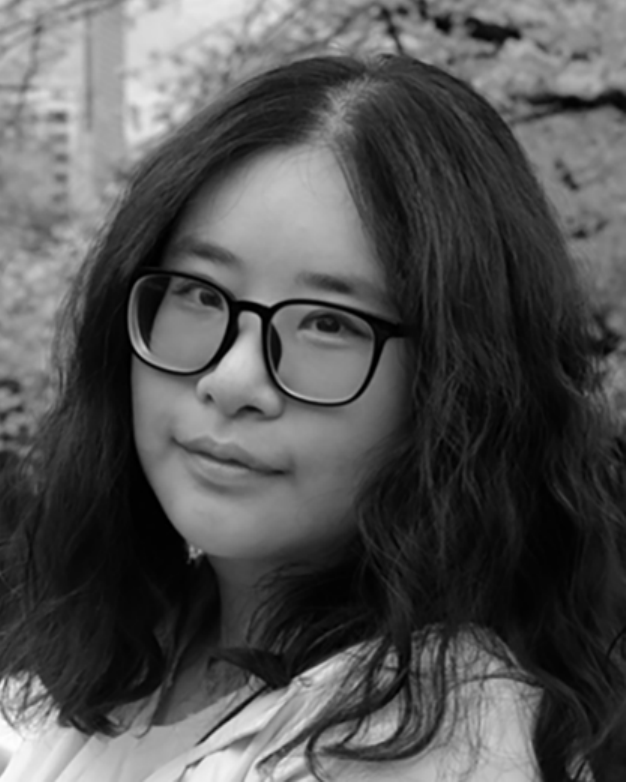}}]{Siyu Wang}
received the M.S. degree in Mechanical Engineering from the University of California, San Diego, in 2018. She is currently pursuing the Ph.D. degree with the Department of Information and Communication Engineering, Institute of Science Tokyo, Yokohama, Japan. Her research focuses on soft body simulations using finite element methods and virtual reality.
\end{IEEEbiography}

\vspace{-32pt} 

\begin{IEEEbiography}[{\includegraphics[width=1in,height=1.25in,clip,keepaspectratio]{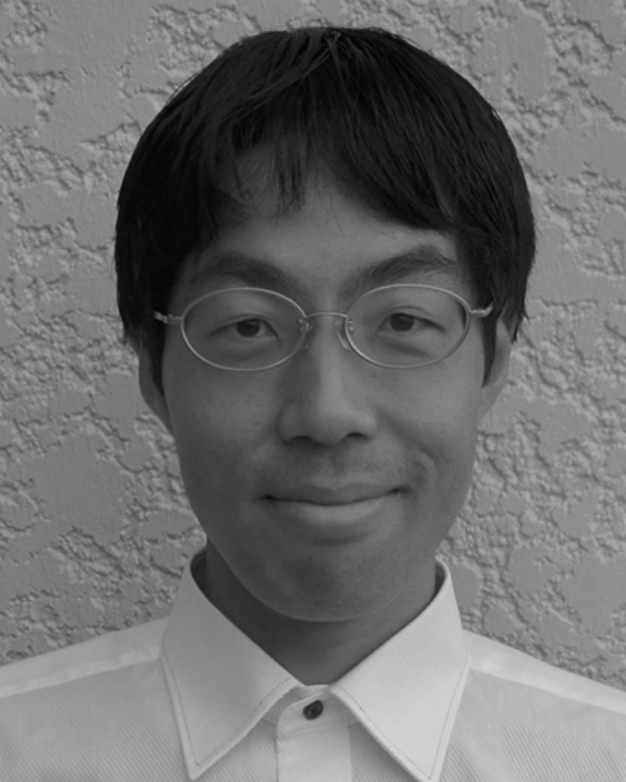}}]{Shoichi Hasegawa}
(Member, IEEE) received the D.Eng. degree in computational intelligence and systems from the Institute of Science Tokyo, Tokyo, Japan. Since 2010, he has been an associate professor with the Institute of Science Tokyo. He was an Associate Professor with the University of Electro Communications, Tokyo. His research interests include haptic renderings, realtime simulations, interactive characters, soft and entertainment robotics, and
virtual reality.
\end{IEEEbiography}

\vspace{11pt}

\vfill

\end{document}